\documentclass[twocolumn,showpacs,amsmath,amssymb,prd]{revtex4}
\def\ba{\begin{eqnarray}}
\def\ea{\end{eqnarray}}

\usepackage{graphicx}
\usepackage{dcolumn}
\usepackage{bm}
\usepackage{epsf}

\begin{document}

\title{Hydrodynamic Simulations in 3+1 General Relativity}

\author{Matthew D. Duez}

\affiliation{Department of Physics, University of Illinois 
	at Urbana-Champaign, Urbana, IL 61801}
	
\author{Thomas W. Baumgarte}
\altaffiliation{Department of Physics, University of Illinois 
	at Urbana-Champaign, Urbana, IL 61801}

\affiliation{Department of Physics and Astronomy, Bowdoin College,
	Brunswick, ME 04011}
 
\author{Pedro Marronetti}
	
\affiliation{Department of Physics, University of Illinois 
	at Urbana-Champaign, Urbana, IL 61801}

\author{Stuart L. Shapiro}
\altaffiliation{Department of Astronomy \& NCSA, 
	University of Illinois at Urbana-Champaign, Urbana, IL 61801}

\affiliation{Department of Physics, University of Illinois 
	at Urbana-Champaign, Urbana, IL 61801}

\begin{abstract}
We solve Einstein's field equations coupled to relativistic
hydrodynamics in full 3+1 general relativity to evolve astrophysical
systems characterized by strong gravitational fields.  We model
rotating, collapsing and binary stars by idealized polytropic
equations of state, with neutron stars as the main application. Our
scheme is based on the BSSN formulation of the field equations.  We
assume adiabatic flow, but allow for the formation of shocks. We
determine the appearance of black holes by means of an apparent
horizon finder. We introduce several new techniques for integrating
the coupled Einstein-hydrodynamics system.  For example, we choose our
fluid variables so that they can be evolved without employing an
artificial atmosphere. We also demonstrate the utility of working in a
rotating coordinate system for some problems.  We use rotating stars
to experiment with several gauge choices for the lapse function and
shift vector, and find some choices to be superior to others.  We
demonstrate the ability of our code to follow a rotating star that
collapses from large radius to a black hole.  Finally, we exploit
rotating coordinates to evolve a
corotating binary neutron star system in a quasi-equilibrium circular
orbit for more than two orbital periods.  

\end{abstract}


\maketitle


\section{Introduction}
\label{intro}

With the availability of unprecedented observational data, the physics
of compact object is entering a particularly exciting phase.  New
instruments, including X-ray and $\gamma$-ray satellites and neutrino
observatories, are detecting signals from highly relativistic events
in regions of strong gravitational fields around neutron stars and
black holes.  A new generation of gravitational wave
interferometers is promising to open a completely new window for the
observation of compact objects.  The ground-based gravity wave
observatories LIGO and TAMA are already operational and are collecting
data, GEO and VIRGO will be completed soon, and a space-based
interferometer LISA is currently under design.

Given the small signal-to-noise ratio in these new gravitational wave
detectors, theoretical models of likely sources are needed for the
positive identification of the signal as well as for its physical
interpretation \cite{c93}.  One promising technique for the identification
of signals in the noise output of the detector is matched filtering,
which requires accurate theoretical gravitational wave templates
\cite{fh98}.  The need for such templates has driven a
surge of interest in developing reliable techniques capable of
their construction.

Compact binaries, i.e.~binaries consisting of either black holes or
neutron stars, are among the most promising sources of gravitational
radiation.  Much progress has been made in refining post-Newtonian point-mass
approximations.  These are suitable for large binary separations for
which relativistic effects are sufficiently small and any
internal structure can be neglected~\cite{bd00}.  At small
binary separations, the most promising technique for modeling the 
inspiral, coalescence and merger is numerical relativity.

Several other observed phenomena involving compact objects require
numerical relativity for their modeling.  One such example is Gamma
Ray Bursts (GRBs).  While it is not yet known what the origin of GRBs
is, the central source is almost certainly a compact
object~\cite{npp92}.  Most scenarios involve a rotating black hole
surrounded by a massive magnetized disk, formed by a supernova, or
the coalescence of binary neutron stars \cite{Piran:2002kw}.  To
confirm or refute any GRB scenario requires numerical studies in full
3+1 relativistic magnetohydrodynamics.

Another astrophysical scenario requiring numerical treatment is the
formation of supermassive black holes (SMBHs).  Among the scenarios
proposed to explain SMBH formation are the collapse of a relativistic
cluster of collisionless matter, like a relativistic star cluster
\cite{rsc} or self-interacting dark matter halo \cite{sidmh}, or
the collapse of a supermassive star~\cite{r01}.
Depending on the details of the collapse, SMBH formation may generate
a strong gravitational wave signal in the frequency band of the
proposed space-based laser interferometer LISA.  Understanding the SMBH
formation route may shed key insight into structure and galaxy formation
in the early universe.

Solving the coupled Einstein field and hydrodynamics equations is a
challenging computational task, requiring the simultaneous solution of
a large number of coupled nonlinear partial differential equations.
In addition to all of the usual problems of numerical hydrodynamics --
handling advection, shock discontinuities, etc -- one encounters the
problems inherent to numerical relativity.  The latter include
identifying a suitable formulation of Einstein's field equations,
enforcing a well-behaved coordinate system, and, if black holes are
formed, dealing with spacetime singularities.  

The construction of self-consistent numerical solutions to the coupled
equations of
relativistic hydrodynamics and gravitation dates back to the pioneering
work of May and White in spherical symmetry \cite{May:1966} (see also
\cite{Font:2000} for a review).  In one of the first attempts to perform
numerical integrations in three spatial dimensions, Wilson, Mathews, and
Marronetti~\cite{Wilson:1989,Wilson:1995ty,Wilson:1996ty} (see
\cite{Flanagan:1999,Mathews:2000} for later corrections) tackled the
binary neutron star problem.  They simplified
Einstein's field equations by assuming that the spatial metric remains
conformally flat at all times.  Their implementation of relativistic
hydrodynamics was based on earlier work by Wilson \cite{Wilson:1972}
and used upwind differencing to handle advection and artificial
viscosity to capture shocks.  The first fully self-consistent
relativistic hydrodynamics code, which treats the gravitational fields
without approximation, was developed by Shibata \cite{Shibata:1999aa}.
This code, based on earlier work by Shibata and Nakamura~\cite{sn95},
adopts a Van Leer hydrodynamics scheme~\cite{vl77,on89} and
also employs artificial viscosity for shocks.  This code has been used
in various astrophysical applications, including the coalescence and
merger of binary neutron stars \cite{Shibata:1999wm,Shibata:2002jb}
and the stability of single, rotating neutron stars
\cite{sbs00a,sbs00b,bss00}.  In an alternative approach,
Font {\it et al}~\cite{fmst00} implemented a more accurate
high-resolution shock-capturing technique to solve the equations of
relativistic hydrodynamics.  This code has been used to study
pulsations of relativistic stars \cite{fgimrssst02}.

In this paper, we report on the status and some astrophysical
applications of our new 3+1 general relativistic hydrodynamics code.
Our code, based on the so-called Baumgarte-Shapiro-Shibata-Nakamura
(BSSN) formulation of Einstein's
equations \cite{sn95,bs98b}, has several novel features, including an
algorithm that does not require the addition of a tenuous, pervasive
atmosphere that is commonly used in Eulerian hydrodynamical codes,
both Newtonian and relativistic.  This ``no atmosphere'' algorithm
proves to be very robust and eliminates many problems associated with
the traditional atmospheric approach \cite{Swesty:1999ke}.

We treat 1D shocks, spherical dust collapse to black holes, and
relativistic spherical equilibrium stars to demonstrate the ability
of our code to accurately evolve the coupled field and
hydrodynamic equations in relativistic scenarios.  We then use the
evolution of stable and unstable uniformly rotating polytropes as a
testbed to determine which gauge conditions are best-behaved in the
presence of strong-field matter sources with significant angular
momentum.  We introduce rotating coordinate systems and show that
these can yield more accurate simulations of rotating objects
than inertial frames.  We demonstrate the ability of our code to
hold accurately stable differentially rotating stars in equilibrium.
We also show that our code can follow
the collapse of rapidly differentially rotating stars reliably until
an apparent horizon appears, by which time the equatorial radius has
decreased from its initial value by more than a factor of ten.

We then turn to simulations of binary neutron stars.  We adopt initial
data describing corotating $n=1$ polytropes in quasi-equilibrium
circular orbit,
and evolve these data for over two orbital periods.  In this paper
we present results for one particular binary and discuss the effect
of corotating frames as well as the outer boundaries.  An extended
study, including binary sequences up to the dynamically identified
innermost stable circular orbit (ISCO), will be presented in a forthcoming
paper \cite{Marronetti:2002}.

This paper is organized as follows.  Secs.~\ref{fields} and
\ref{hydro} describe our method of evolving the field and hydrodynamic
equations, respectively.  Sec.~\ref{gauge_choices} summarizes the
various gauge choices with which we experiment.  Sec.~\ref{diag} lists
the diagnostics used to gauge the reliability of our simulations.
Sec.~\ref{tests} describes several tests of our algorithm.
Sec.~\ref{iso_stars} applies our formalism to evolve non-rotating,
uniformly rotating, and differentially rotating polytropes.  In
Sec.~\ref{Bin} sketches our binary neutron star calculations.  Our
results are summarized in Sec. \ref{summary}.  Some details of our
hydrodynamic scheme and the rotating frame formalism are presented in
the appendices.


\section{Gravitational Field Evolution}
\label{fields}

\subsection{Basic Equations}
\label{field_eqs}

We write the metric in the form
\begin{equation}
ds^2 = -\alpha^2 dt^2 + \gamma_{ij}(dx^i+\beta^idt)(dx^j+\beta^jdt),
\end{equation}
where $\alpha$, $\beta^i$, and $\gamma_{ij}$ are the lapse, shift, and
spatial metric, respectively.  The extrinsic curvature $K_{ij}$ is
defined by
\begin{equation}
\label{Kij}
(\partial_t - {\mathcal{L}}_{\beta})\gamma_{ij} = -2\alpha K_{ij},
\end{equation}
where ${\mathcal{L}}_{\beta}$ is the Lie derivative with respect to
$\beta^i$.  We choose geometrized units with $G = c = 1$ throughout,
so Einstein's field equations are
\begin{equation}
\label{Einstein}
G_{\mu\nu} = 8\pi T_{\mu\nu}.
\end{equation}
We use greek letters to denote spacetime indices, and latin letters
for spatial indices.  Using the above variables, the field equations
(\ref{Einstein})
split into the usual 3+1 ADM equations \cite{adm62}.  These consist of
the Hamiltonian constraint
\begin{equation}
\label{Hamiltonian_ADM}
R - K_{ij}K^{ij} + K^2 = 16\pi\rho,
\end{equation}
the momentum constraint
\begin{equation}
\label{momentum_ADM}
D_jK^j_i - D_iK = 8\pi S_i,
\end{equation}
and the evolution equation for $K_{ij}$
\begin{eqnarray}
\label{evolve_ADM}
(\partial_tK_{ij} - {\mathcal{L}}_{\beta}K_{ij}) &=&
        - D_iD_j\alpha \nonumber \\
    & & + \alpha(R_{ij} - 2K_{il}K^l{}_j
        + K K_{ij} \\
	& &\qquad - 8\pi(S_{ij}+{1\over 2}\gamma_{ij}(\rho-S^i{}_i)))
	\nonumber
\end{eqnarray}
in addition to (\ref{Kij}).  Here $D$, $R_{ij}$ and $R$ are the
covariant derivative operator, the three-dimensional Ricci tensor and the
scalar curvature associated with $\gamma_{ij}$.  The matter source
terms $\rho$, $S_i$, and $S_{ij}$ are projections of the stress-energy
tensor with respect to the unit normal $n^{\alpha}$ on the time slice
\begin{eqnarray}
  \label{rho_def}
  \rho &=& n_{\alpha}n_{\beta}T^{\alpha\beta} \nonumber \\
  S_i  &=& -\gamma_{i\alpha}n_{\beta}T^{\alpha\beta} \nonumber \\
  S_{ij} &=& \gamma_{i\alpha}\gamma_{j\beta}T^{\alpha\beta}. 
\end{eqnarray}
Since numerical implementations of the ADM equations typically develop
instabilities after very short times, we use a reformulation of these
equations that is now often referred to as the BSSN
formulation~\cite{sn95,bs98b}.  This reformulation consists of
evolving the conformally related metric $\tilde\gamma_{ij}$, the
conformal exponent $\phi$, the trace of the extrinsic curvature $K$,
the conformal traceless extrinsic curvature $\tilde A_{ij}$, and the
conformal connection functions $\tilde\Gamma^i$ defined by
\begin{eqnarray}
\label{YLsplit}
\gamma_{ij} &=& e^{4\phi}\tilde\gamma_{ij} \\
K_{ij} &=& e^{4\phi}\bigl(\tilde A_{ij} + {1\over 3}\tilde\gamma_{ij}K\bigr) \\
\tilde\Gamma^i &=& -\tilde\gamma^{ij}{}_{,j},
\end{eqnarray}
where $\det(\tilde\gamma_{ij}) = 1$ and $\mbox{tr}(\tilde A_{ij}) =
0$.  In terms of these variables, Eqs.~(\ref{Kij}) and
(\ref{evolve_ADM}) become
\begin{eqnarray}
\label{evolve_gamma}
(\partial_t - {\mathcal{L}}_{\beta})\tilde\gamma_{ij}
		&=& -2\alpha\tilde A_{ij} \\
\label{evolve_phi}
(\partial_t - {\mathcal{L}}_{\beta})\phi
                &=& -{1\over 6}\alpha K \\
\label{evolve_K}
(\partial_t - {\mathcal{L}}_{\beta})K
                &=& -\gamma^{ij}D_jD_i\alpha + {1\over 3}\alpha K^2 \\
                & & + \alpha \tilde A_{ij}\tilde A^{ij}
                    + 4\pi\alpha (\rho + S) \nonumber \\
\label{evolve_A}
(\partial_t - {\mathcal{L}}_{\beta})\tilde A_{ij}
                &=& e^{-4\phi}(-D_iD_j\alpha
                    + \alpha(R_{ij}-8\pi S_{ij}))^{TF} \nonumber \\
		& & + \alpha(K\tilde A_{ij} - 2\tilde A_{il}\tilde A^l{}_j)
\end{eqnarray}
and
\begin{eqnarray}
\label{evolve_Gamma}
\partial_t\tilde\Gamma^i &=& \partial_j(2\alpha\tilde A^{ij} 
		+ {\mathcal{L}}_{\beta}\tilde\gamma^{ij}) \nonumber \\
	&=& \tilde\gamma^{jk}\beta^i{}_{,jk} 
		+ {1\over 3}\tilde\gamma^{ij}\beta^k{}_{,kj}
 - \tilde\Gamma^j\beta^i{}_{,j} \\
 & &+{2\over 3}\tilde\Gamma^i\beta^j{}_{,j} 
    + \beta^j\tilde\Gamma^i{}_{,j} - 2\tilde A^{ij}\partial_j\alpha 
	\nonumber \\
 & &- 2\alpha\left({2\over 3}\tilde\gamma^{ij}K_{,j} - 6\tilde A^{ij}\phi_{,j} 
    - \tilde\Gamma^i{}_{jk}\tilde A^{jk} + 8\pi\tilde\gamma^{ij}S_j
	\right)
	\nonumber
\end{eqnarray}
(see \cite{bs98b} for the computation of the Lie derivatives.)

In terms of the BSSN variables, the constraint equations
(\ref{Hamiltonian_ADM}) and (\ref{momentum_ADM}) become,
respectively,
\begin{eqnarray}
\label{Hamiltonian_BSSN}
  0 = \mathcal{H} &=& 
                \tilde\gamma^{ij}\tilde D_i\tilde D_j e^{\phi}
                - {e^{\phi} \over 8}\tilde R  \\
		& & + {e^{5\phi}\over 8}\tilde A_{ij}\tilde A^{ij}
		    - {e^{5\phi}\over 12}K^2 + 2\pi e^{5\phi}\rho, 
	\nonumber \\
\label{momentum_BSSN}
  0 = {\mathcal{M}}^i &=&
  \tilde D_j(e^{6\phi}\tilde A^{ji})- {2\over 3}e^{6\phi}\tilde D^i K
  - 8\pi e^{6\phi}S^i,
\end{eqnarray}
where $S^i = \tilde\gamma^{ij}S_j$. 
While the two constraints are identically zero for analytical
solutions, they vanish only approximately in numerical calculations.
Thus, the Hamiltonian and momentum constraint residuals $\mathcal{H}$
and $\mathcal{M}$ can be monitored as a code test during numerical
evolution calculations.  In the BSSN formulation, we also monitor the
new constraint
\begin{equation} \label{gamma_BSSN}
0 = \mathcal{G}^i = \tilde \Gamma^i + \tilde \gamma^{ij}_{~~,j}.
\end{equation}


\subsection{Boundary Conditions}
\label{BC_analytic}

Like any other hyperbolic system, the Einstein field equations must be
supplemented by initial conditions and boundary conditions to have a
unique evolution.  We adopt boundary conditions that follow from the
assumption of asymptotic flatness, i.e.  $g_{\alpha\beta}\rightarrow
\eta_{\alpha\beta}$. In the asymptotic domain, monopole terms dominate
in the longitudinal variables, so $\phi \propto r^{-1}$.  The
transverse fields will be dominated by outgoing gravitational waves,
so $\tilde\gamma_{ij}-\eta_{ij} \propto f(t-r)r^{-1}$ and $\tilde
A_{ij} \propto g(t-r)r^{-1}$, where $f$ and $g$ are unknown functions
of retarded time.  Note that $r^{-1}$ is a special case of
$f(t-r)r^{-1}$, so that $\phi$, $\tilde\gamma_{ij}$, and $\tilde
A_{ij}$ all satisfy outgoing wave boundary conditions.  The
appropriate boundary conditions for $K$ and $\tilde\Gamma^i$ depend on
the gauge conditions used in the interior.


\subsection{Numerical Implementation}
\label{field_implementation}

We evolve Eqs.~(\ref{evolve_gamma})-(\ref{evolve_Gamma}) using an
iterative Crank-Nicholson scheme with one predictor step and two
corrector steps \cite{t00}.  In this algorithm a function $f$ with
time derivative $\dot f$ is updated from its value $f^n$ at timestep
$n$ to its value $f^{n+1}$ at the next timestep $n+1$ a time $\Delta
T$ later.  In the explicit predictor step ${}^1f^{n+1} = f^n + \Delta
T \dot f^n$, where $\dot f^n$ is computed from quantities on timestep
$n$, a ``predicted'' new value ${}^1f^{n+1}$ is found.  In the
following two corrector steps, ${}^{2}f^{n+1} = f^n + \Delta T(\dot f^n
+ {}^1\dot f^{n+1})/2$ and $f^{n+1} = {}^{3}f^{n+1} = f^n + \Delta
T(\dot f^n + {}^2\dot f^{n+1})/2$, these predicted values are
``corrected''.  The final value $f^{n+1}$ converges quadratically in
$\Delta T$.  $\Delta T$ is set by the Courant factor: 
$C = \Delta T/\Delta x$, where $\Delta x$ is the
coordinate distance between adjacent gridpoints.  We typically use
$C = 0.5$.  The code implementing this evolution scheme
has been discussed elsewhere \cite{bs98b}, so we will highlight here
only the new features of our code.

We enforce the algebraic constraints $\det{\tilde\gamma_{ij}} = 1$ and
$\mbox{tr}(\tilde A_{ij}) = 0$ as described in~\cite{yo02}.  Also
following \cite{yo02}, we replace the term ${2\over
3}\tilde\Gamma^i\beta^j{}_{,j}$ in Eq.~(\ref{evolve_Gamma}) with the
analytically equivalent $-(\tilde\gamma^{ij}{}_{,j} + {1\over
3}\tilde\Gamma^i)\beta^j{}_{,j}$.  These changes have little effect on
the evolutions described in this paper, but lead to significant
improvements when treating black holes by excision boundary conditions
\cite{yo02}.

We use second order centered differencing for all spatial derivatives
in the field equations.  We have not found it necessary to use upwind
differencing for any derivatives.  We did find, however, that the
addition of some dissipation in the evolution equation for $\phi$
increases the stability of the code.  This can be supplied by upwind
differencing of the term which advects $\phi$ along the shift, and we
have confirmed that this will indeed improve the stability. 
However, we have chosen instead to add the Hamiltonian
constraint to the evolution equation for $\phi$, as follows
\begin{equation}
\label{phi_nevolve}
(\partial_t - {\mathcal{L}}_{\beta})\phi
                = -{1\over 6}\alpha K + c_H \mathcal{H}.
\end{equation}
Here the parameter $c_H$ is set between $0.02\Delta T$ and $0.06\Delta T$. 
$c_H \mathcal{H}$ is a diffusive term, with Courant condition given
\cite{nr} by $2c_H\Delta T/(\Delta x)^2 \leq 1$, so making $c_H$
proportional to $\Delta T$ is necessary in order to avoid an instability
at high resolutions.  It also provides dimensional consistency in
(\ref{phi_nevolve}).  Using Eq.~(\ref{phi_nevolve}) offers the advantage
of significantly decreasing the growth in the error of the Hamiltonian
constraint (see Sec.~\ref{Bin} for an example of this.)  We note that
the above is similar to one of the modifications of BSSN suggested in
\cite{ys02}.


\subsection{Implementation of Boundary Conditions}
\label{fields_BCs}

As discussed in Section (\ref{BC_analytic}), we use Sommerfeld
boundary conditions for most of the field variables.  That is, the
value of a quantity $f$ on the boundary at time $t$ and distance $r$
from the origin is
\begin{equation}
\label{wavelike}
f(r,t) = {r - \Delta r\over r}f(r-\Delta r, t-\Delta T),
\end{equation}
where $\Delta T$ is the timestep and $\Delta r = \alpha
e^{-2\phi}\Delta T$.  

For the functions $\tilde\Gamma^i$ we have experimented with several
boundary conditions.  We find little sensitivity to the condition
used; the best choice seems to be fixing $\tilde\Gamma^i$ at their
initial values (zero, for most of the applications here).


\section{Relativistic Hydrodynamics}
\label{hydro}

\subsection{Basic Equations}
\label{hydro_equations}
We describe the matter source of the Einstein equations as a perfect fluid
so that the stress-energy tensor can be written
\begin{equation}
\label{perfect_fluid}
T_{\mu\nu} = (\rho_0 + \rho_0\epsilon + P)u_{\mu}u_{\nu} + P g_{\mu\nu}.
\end{equation}
Here $\rho_0$, $\epsilon$, $P$, and $u_{\mu}$ are the rest-mass
density, specific internal energy, pressure, and fluid four-velocity,
respectively.  We adopt a
$\Gamma$-law equation of state
\begin{equation}
\label{ideal_P}
P = (\Gamma - 1)\rho_0\epsilon,
\end{equation}
where $\Gamma$ is a constant.  For isentropic flow,
this is equivalent to the polytropic relation
\begin{equation}
\label{polytrope}
P = \kappa\rho_0^{\Gamma}\ ,
\end{equation}
where $\kappa$ is a constant.  In our simulations we encounter
non-isentropic flow (due to shocks), and hence use equation
(\ref{ideal_P}).

The equations of motion follow from
the continuity equation
\begin{equation}
\label{continuity}
\nabla_{\mu}(\rho_0 u^{\mu}) = 0
\end{equation}
and the conservation of stress-energy
\begin{equation}
\label{eom}
T^{\mu\nu}{}_{;\nu} = 0\ .
\end{equation}
Following~\cite{Shibata:1999aa}, these equations can be brought into
the form
\begin{eqnarray}
\label{evolve_rhostar}
\partial_t\rho_{\star} + \partial_i(\rho_{\star}v^i) &=& 0 \\
\label{evolve_estar}
\partial_t e_{\star} + \partial_i(e_{\star}v^i) &=& 0 \\
\label{evolve_stilde}
\partial_t\tilde S_k + \partial_i(\tilde S_k v^i) &=& 
	-\alpha e^{6\phi}P_{,k} - wh\alpha_{,k} \\
	& & - \tilde S_j\beta^j{}_{,k} + {\alpha e^{-4\phi}\tilde S_i
	\tilde S_j\over 2wh}\tilde\gamma^{ij}{}_{,k} \nonumber \\
  & & \displaystyle - {2\alpha h(w^2-\rho_{\star}{}^2)\over w}\phi_{,k}\ ,
	\nonumber
\end{eqnarray}
where $h = 1+\epsilon+P/\rho_0$, $\rho_{\star} = \rho_0\alpha u^0
e^{6\phi}$, $w = \rho_{\star}\alpha u^0$,
$e_{\star}=(\rho_0\epsilon)^{1/\Gamma}\alpha u^0e^{6\phi}$, $\tilde
S_k = \rho_{\star}hu_k$, and $v^i = u^i/u^0$ is the 3-velocity.  The
quantity $w$ is determined by the normalization condition
$u^{\nu}u_{\nu} = -1$, which can be written
\begin{equation}
\label{w}
\displaystyle w^2 = \rho_{\star}^2 
        + e^{-4\phi}\tilde\gamma^{ij}\tilde S_i \tilde S_j
	\left[1 + {\Gamma e_{\star}{}^{\Gamma}\over
	\rho_{\star}(we^{6\phi}/\rho_{\star})^{\Gamma-1}}\right]^{-2}\ .
\end{equation}

The perfect fluid given by Eq.~(\ref{perfect_fluid}) generates the following
source terms for the ADM equations
\begin{eqnarray}
\label{fluid_sources}
\rho &=& h w e^{-6\phi} - P \\
S_i &=& e^{-6\phi} \tilde S_i \\
\displaystyle S_{ij} &=& {e^{-6\phi}\over w h}\tilde S_i \tilde S_j
                          + P \gamma_{ij}
\end{eqnarray}

We will only be considering systems where there is vacuum everywhere
outside the star or stars.  Therefore, the appropriate boundary
condition on the matter flow is that no material should be flowing
into the grid through the outer boundaries.


\subsection{Numerical Implementation}
\label{hydro_avd}

We evolve the hydrodynamic variables using an iterative
Crank-Nicholson scheme.  This scheme is slightly different from the
one used to update the field variables.  In the corrector steps,
instead of weighting $\dot f^n$ and ${}^i \dot f^{n+1}$ equally
(i.e.~${}^{i+1}f^{n+1} = f^n + \Delta T(0.5\dot f^n + 0.5\ {}^i\dot
f^{n+1})$), we make the evolution more implicit by setting
${}^{i+1}f^{n+1} = f^n + \Delta T(0.4\dot f^n + 0.6\ {}^i\dot
f^{n+1})$.  This makes the code slightly more stable.  

As is often done in hydrodynamics codes~\cite{sn92}, the updating of
the fluid variables onto a new timestep is divided into two steps
(``operator splitting''): the advection step (accounting for the
advective terms on the left-hand sides of
Eqs.~(\ref{evolve_rhostar})-(\ref{evolve_stilde})), and the source
step (accounting for the right-hand sides of
Eqs.~(\ref{evolve_rhostar})-(\ref{evolve_stilde})).  Each step of a
Crank-Nicholson update consists of applying first an advection substep
and then a source substep.  Our scheme for carrying out the advection
substep is similar to the Van Leer scheme, and is discussed in detail
in Appendix \ref{advection_appendix}.  Since
Eq.~(\ref{evolve_rhostar}) has no sources, $\rho_{\star}$ is
completely updated after it is advected.
Following~\cite{Swesty:1999ke}, we then use the updated $\rho_{\star}$
to complete the updating of $e_{\star}$ and $\tilde S_k$.  It is shown
in \cite{Swesty:1999ke} that this gives improved behavior in Newtonian
simulations of binary polytropes.


\subsection{Artificial Viscosity}
\label{art_vis}

We handle shocks by adding quadratic artificial viscosity. This
consists of adding a viscous pressure \cite{Shibata:1999aa}
\begin{equation}
\label{art_visc}
P_{\rm Qvis} = \left\{ \begin{array}{ll} \displaystyle 
                C_{\rm Qvis} A (\delta v)^2 
		& \mbox{for $\delta v<0$}\ , \\
		0 & \mbox{otherwise}\ ,
		\end{array} \right.
\end{equation}
where $A$ is defined as
${e_{\star}^{\Gamma} \over (we^{6\phi}/\rho_{\star})^{\Gamma-1}}$ and
$\delta v = 2\partial_k v^k\Delta x$.  Shock heating causes an
increase in the local internal energy.  Following~\cite{Shibata:1999aa},
we change equation (\ref{evolve_estar}) to
\begin{equation}
\label{shock_heating}
\partial_t e_{\star} + \partial_i(e_{\star}v^i)
  = -(\rho_0\epsilon)^{-1+1/\Gamma} {P_{\rm Qvis}\over\Gamma}
  \partial_k\left({we^{6\phi}v^k\over\rho_{\star}}\right)
\end{equation}

We have also implemented linear artificial viscosity terms \cite{mw93}
that can be used to dissipate radial oscillations triggered in stars
by the truncation error associated with finite differencing. The
corresponding addition to the pressure is
\begin{eqnarray}
P_{\rm Lvis} = \left\{ \begin{array}{ll} \displaystyle 
                -C_{\rm Lvis} \sqrt{(\Gamma/n) \rho_{\star} A} ~\delta v 
		& \mbox{for $\delta v<0$}\ , \\
		0 & \mbox{otherwise}\ .
		\end{array} \right.
\end{eqnarray}

\begin{figure}
\epsfxsize=2.8in
\begin{center}
\leavevmode \epsffile{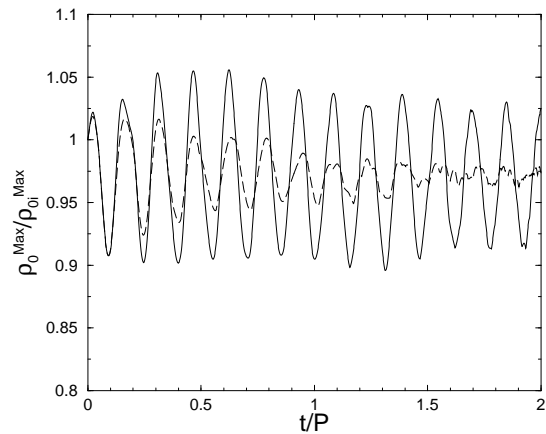}
\end{center}
\caption{ Maximum rest mass density $\rho_0$ as a fraction of its initial
  value $\rho_{0i}$ for the binary system shown in figure
  \ref{bin_contour}. Stellar radial oscillations can be efficiently quenched
  by the proper use of linear viscosity, as shown here. The solid line
  shows the evolution without linear artificial viscosity, while the dashed
  line shows the effect of this dissipative term.}
\label{bin_lin_vis}
\end{figure}

Linear viscosity can be used at the beginning of a run to drive the
initial data to dynamical equilibrium and later switched off.  Figure
\ref{bin_lin_vis} shows an example of how the radial oscillations can
be quenched by linear viscosity. For this particular example, the
$P_{\rm Lvis}$ was active only where the rest mass density exceeded a
particular threshold value, to force this dissipative effect only deep
inside the neutrons stars.  The (small) dissipated kinetic energy goes
into thermal energy.  We typically use $0.1 \leq C_{\rm Qvis} \leq 1.0$. 
Linear artificial viscosity is not used in the runs described below.

\subsection{Non-Atmospheric Hydrodynamics}
\label{no_atmos}

Numerical work in Eulerian hydrodynamics, both Newtonian and
relativistic, has typically required the presence of a pervasive tenuous
``atmosphere'' that covers the computational grid outside the stars.
To our knowledge, most published codes to date need to keep a
minimum nonzero density that is usually set to be several orders of
magnitude smaller than the maximum stellar density.  Such an
atmosphere has been necessary to prevent overflows arising from
dividing by density in cells devoid of matter.  This artificial
atmosphere has to be small enough not to affect the true dynamical
behavior of the system. However, very small values will propagate the
round-off numerical error very quickly every time a division by the
density is performed. A problem with the presence of this atmosphere
is that as soon as the time evolution starts the material begins to
fall onto the star, creating accretion shocks. Swesty {\it et al.}
\cite{Swesty:1999ke} solved this problem by adding a non-zero
temperature to the atmosphere to restore some sort of equilibrium that
would counterbalance the infall. Also, in order to avoid the bow
shocks generated in the atmosphere by two stars in circular orbital
motion, these authors provide the atmosphere with initial angular
velocity. These are some of the typical problems present in the
traditional artificial atmosphere approach found in many Eulerian
hydrodynamics schemes.

In this paper we present a very simple algorithm that does not require
the presence of atmospheric material. It consists of two ingredients.
The first is the use of the spatial components of the linear momentum
variable $S_k$ as our hydrodynamical variable \cite{note1} instead of
the traditional fluid four-velocity spatial components $u^i$ used in 
most hydrodynamical codes (see for instance \cite{Shibata:1999aa}). 
In the latter case, the Euler equation is used to update the flux
$(\rho_\star ~u^i)$.  Once this update is completed, the dynamical
field $u^i$ is recovered by dividing by the density
$\rho_{\star}$. Using $S_k$ as a variable, we avoid these
divisions. The only time when the variable $u^i$ needs to be
calculated explicitly is when we need the three-velocity $v^i$ that
appears in every advection term on the left hand side of
Eqs. (\ref{evolve_rhostar}-\ref{evolve_stilde}). To avoid doing this
calculation for very low values $\rho_{\star}$, we add the second 
ingredient: the introduction of a threshold value $\rho_{\star~min}$ 
below which all the hydrodynamical fields are set to vacuum values 
(i. e. $\rho_{\star}=u^i=v^i=0$). A typical value for $\rho_{\star~min}$ 
is $10^{-7}$ times the maximum initial value of $\rho_{\star}$.

However, as the time evolution progresses, a tenuous shell of material
typically drifts away from the stars and creates regions of very low density
outside our stars.  If nothing special is done about them,
small shocks will heat this low-density region to very high
temperatures and will generate large velocities.  Although this
low-density region has a negligible effect on our stars and
spacetimes, it can cause the code to crash.  Therefore, we impose a
heating limit outside the star
\begin{equation}
\label{elimit}
e_{\star} = \min(e_{\star},10\rho_{\star})
\mbox{~~~if~~~} \rho_{\star}< e_{\rm factor} \times \rho_{\star \rm max}, 
\end{equation}
where $e_{\rm factor}$ is a constant that is determined empirically
for a given physical scenario.  We generally choose values between
$10^{-3}$ and $10^{-6}$, where the larger values of $e_{\rm factor}$
were only needed in simulations of collapsing stars with a strong
bounce.  We note that this is similar to the technique used
in~\cite{fmst00}, in which the polytropic equation of state
(\ref{polytrope}) is applied in the low-density region outside the
star or stars.


\subsection{Boundary Conditions}
\label{hydro_BCs}

Since matter often diffuses outward, albeit in minute quantities, from the
surface of the star(s) to the boundaries, we need to impose boundary
conditions on the matter at the outer grid points.  In algorithms
where an artificial atmosphere is present, it is crucial to choose
boundary conditions which do not lead to a continuous inflow from the
boundary, or to bad behavior in the atmosphere.  By eliminating such
an atmosphere, however, all reasonable boundary conditions yield the
same behavior so long as the boundaries are placed far enough from the
star(s) that little matter ever reaches them.

We usually use an outflow boundary condition. For example, if the
$x$-coordinate of gridpoints is indexed by an integer $i$ with
$i_{\rm min}\leq i \leq i_{\rm max}$, this boundary condition at the
outer-$x$ boundary $i = i_{\rm max}$ is implemented as
\begin{eqnarray}
\label{r_bc}
\rho^{n+1}_{\star \rm imax} &=& \rho^{n+1}_{\star \rm imax-1} \\
\label{e_bc}
e^{n+1}_{\star \rm imax} &=& e^{n+1}_{\star \rm imax-1} \\
\label{s_bc}
\tilde S^{n+1}_{\rm imax} &=&  \left\{ \begin{array}{ll}
                           \tilde S^{n+1}_{\rm imax-1} 
			           \mbox{if $\tilde S^{n+1}_{\rm imax-1}>0$}\\
			   0 & \mbox{otherwise}
			   \end{array}\right.
\end{eqnarray}

We have experimented with other boundary conditions as well.  We have tried
fixing $\rho_{\star}$, $e_{\star}$, and $\tilde S_k$ at their initial values. 
We have also tried simply copying the adjacent gridpoint onto the boundary
with no outflow restrictions ({\it Copy}). These conditions produce similar 
results to those of the outflow condition for all applications, while being
somewhat less computationally expensive.


\section{Gauge Choices}
\label{gauge_choices}

\subsection{Lapse}
\label{gauge_lapse}

We experiment with several time slicing conditions.  First, we try
maximal slicing, which enforces $K = \partial_tK = 0$
\begin{equation}
\label{maximal_slicing}
\mbox{mx:}\quad 0 = -\gamma^{ij}D_jD_i\alpha + \alpha
        \tilde A_{ij}\tilde A^{ij}
	+ 4\pi\alpha(\rho + S)\ .
\end{equation}
This slicing condition has the advantages of controlling $K$ and
avoiding singularities.  Unfortunately, it is a computationally
expensive gauge choice, since it involves solving an elliptic PDE
every timestep.  Therefore, we also try a slicing condition which
approximates maximal slicing, the so-called ``K-driver'' proposed by
Balakrishna {\it et al}~\cite{betal96}.  The idea is to convert the
elliptic equation (maximal slicing) into a parabolic evolution
equation
\begin{equation}
\label{K_driver}
\mbox{Kdr:}\quad
\partial_t\alpha = -\epsilon(\partial_tK + cK)\ ,
\end{equation}
where $\epsilon$ and $c$ are positive constants.  The equation $\partial_tK =
-cK$, corresponding to exponential decay in $K$, is the solution of
equation (\ref{K_driver}) as $\epsilon \rightarrow \infty$.  However,
setting $\epsilon$ at too large a value in our code will produce a
numerical instability.  (See the discussion of $c_H$ in Sec.
\ref{field_implementation}.) Fortunately, this limitation can be overcome.
We are able to effectively evolve with larger $\epsilon$ by breaking up
each timestep into several substeps and evolve Eq.(\ref{K_driver}) using
a smaller $\Delta T$ than that used by the other variables.  On each
substep, we use the values of the metric on the destination time level, so
the process is equivalent to solving the elliptic equation
$\partial_tK + cK = 0$ by relaxation, except that we do not carry the
process to convergence.  Instead, we typically use 5 substeps per step,
with $\epsilon = 0.625$ and $c = 0.1$.

An even less computationally expensive lapse condition is harmonic
slicing, which for vanishing shift reduces to
\begin{equation}
\label{harmonic}
\mbox{hm:}\quad
\partial_t(\alpha^{-1}\gamma^{-1/2}) = 0~.
\end{equation}
We apply this condition unchanged for vanishing and non-zero shift,
and find that it often gives behavior similar to that obtained by
using the above two slicings.


\subsection{Shift}
\label{gauge_shift}

We also experiment with different spatial gauge choices.  The simplest
admissible shift choice, which turns out to be surprisingly good for
collapsing star applications, is to keep the shift ``frozen'' at its
initial values
\begin{equation}
\label{frozen}
\mbox{fz:} \quad
\beta^i(t) = \beta^i(0)  
\end{equation}
at each grid point.

We also try the approximate minimal distortion (AMD) gauge introduced by
Shibata~\cite{s99}
\begin{equation}
\label{amd}
\mbox{AMD:}\quad
\delta_{ij}\nabla^2\beta^i + {1\over 3}\beta^k{}_{,kj} = J_j \,
\end{equation}
where $\nabla^2$ is the flat-space Laplacian and
\begin{equation}
\label{amd_source}
J_i = 16\pi\alpha S_i + 2\tilde A_{ij}(\alpha^{,j}-6\alpha \phi^{,j})
      + {4\over 3}\alpha K_{,i} .
\end{equation}
This gauge condition was designed to approximate the Smarr and York
minimal distortion shift condition \cite{sy78}, which in turn was
constructed to minimize gauge-related time variation in the spatial
metric.

As Shibata points out~\cite{sbs00a}, the AMD condition must be
modified in the event of a collapse in order to prevent a ``blowing
out'' of coordinates on the black hole throat, which manifests itself
by a growth in the proper 3-volume element, i.e. by growth in
$\phi$.  The blowing out can be controlled by preventing the radial
component of the shift from becoming large and positive
\begin{equation}
\label{mamd}
\mbox{MAMD:}\quad
\beta^i = \left\{ \begin{array}{ll}
                \beta^i_{\rm AMD} 
		& \mbox{for $\phi_c < (4/3)\phi_{c i}$} \\
\displaystyle	\beta^i_{\rm AMD} - f\beta^r_{\rm AMD}{x^i\over r}
		  & \mbox{otherwise}
		\end{array} \right.,
\end{equation}
where $\beta^i_{AMD}$ is the solution of equation (\ref{amd}),
$\phi_c$ is the value of $\phi$ at the coordinate origin, $\phi_{c i}=
\phi_c(t=0)$, and
\begin{eqnarray}
\label{AMD_supp}
f &=& \left({3\phi_c\over 2\phi_{c i}} - 2\right){1\over 1+(r/R)^4} \\
\beta^r_{\rm AMD} &=& x^k\beta^k_{\rm AMD}/r,
\end{eqnarray}
where $R$ is a constant.  This correction is only useful in
configurations with near spherical symmetry, so that the collapse is
nearly radial at the center.  It is disabled for simulations of binary
systems.

Finally, we try approximating the ``Gamma-freezing'' condition
$\partial_t\tilde\Gamma^i = 0$ using a ``Gamma-driver'', which
enforces controls $\tilde\Gamma^i$ in the same way that the K-driver
controls $K$
\begin{equation}
\label{gamma_driver}
\mbox{Gdr:}\quad
\partial_t\beta^i = k(\partial_t\tilde\Gamma^i + \eta\tilde\Gamma^i) .
\end{equation}
Here $k$ and $\eta$ are positive constants, and, as with the K-driver,
we can effectively make $k$ larger than would otherwise be possible
by breaking up each step into multiple substeps.  This
shift condition has been used successfully in black hole evolution
calculations~\cite{ab01}.  The Gamma-freezing condition is closely
related to minimal distortion (and hence approximate minimal
distortion), and it is hence not surprising that the modification
(\ref{mamd}) must also be applied to the Gamma-driver shift.  Typical
values for the Gamma-driver's parameters are $k=0.01$ and $\eta=0.2$,
using 10 substeps per step.

We have also implemented of the full Gamma freezing condition
($\partial_t\tilde\Gamma^i = 0$).  However, applying this
condition requires solving three coupled elliptic equations each
timestep (see Eq. (\ref{evolve_Gamma})), and we have found Gamma freezing
to be too computationally expensive to be worth solving exactly.


\subsection{Rotating Frames}
\label{rot_frames}

Rotating coordinate frames possess superior angular momentum
conservation capability over inertial frames in many applications,
such as the hydrodynamical evolution of binaries systems.  In
transforming from an inertial frame with coordinates
$(\bar{t},\bar{x},\bar{y},\bar{z})$ to a rotating frame with
coordinates $(t,x,y,z)$ and constant angular frequency
$\vec{\Omega}=\Omega \vec{e_z}$, we apply the following relations
\begin{eqnarray}
\bar{t} & = & t \nonumber \\
\bar{x} & = & x ~ \cos(\Omega t) - y ~ \sin(\Omega t) \nonumber \\
\bar{y} & = & x ~ \sin(\Omega t) + y ~ \cos(\Omega t) \nonumber \\
\bar{z} & = & z ,
\label{transf_law}
\end{eqnarray}
where the barred variables will represent quantities in the inertial
frame in the remainder of this section.  It is convenient to compare
variables in the two frames at an instant $\bar{t}=t=0$ at which
the two frames are aligned.  At this instant, the line
element transforms from
\begin{eqnarray}
d\bar{s}^2 = -(\bar{\alpha}-\bar{\beta}^i \bar{\beta}_{i}) d\bar{t}^2 
+ 2 \bar{\beta}_{i} d\bar{x}^i d\bar{t}+\bar{\gamma}_{ij} d\bar{x}^i 
d\bar{x}^j \nonumber 
\end{eqnarray}
to
\begin{eqnarray}
ds^2 & = & -\left( \bar{\alpha} -\bar{\gamma}_{ij} 
(\bar{\beta}^i + (\vec{\Omega} \times \vec{r})^i)
(\bar{\beta}^j + (\vec{\Omega} \times \vec{r})^j) \right) dt^2    \nonumber \\
& + & 2 \bar{\gamma}_{ij} (\bar{\beta}^i + (\vec{\Omega} \times \vec{r})^i) 
dx^j dt + \bar{\gamma}_{ij} dx^i dx^j, \nonumber
\end{eqnarray}
where $\vec{r} \equiv (x,y,z)$.
From this equation, we see that the following transformation rules apply
at $\bar{t}=t=0$:
\begin{eqnarray}
\alpha & = & \bar{\alpha} \nonumber \\
\beta^i & = & \bar{\beta}^i + (\vec{\Omega} \times \vec{r})^i \nonumber \\
\gamma_{ij} & = & \bar{\gamma}_{ij},     
\label{rot_fields}
\end{eqnarray}
Eq.~(\ref{rot_fields}) provides the transformation rules for the initial
metric data from
the inertial frame (where it is usually derived) to the rotating frame. 
The only change is the addition of a new term in the shift.  At later
times, vectors and tensors in the two frames will also differ by a
rotation.  However, we note that at all times there will be some inertial
frame, related to $(\bar{t},\bar{x},\bar{y},\bar{z})$ by a rotation matrix,
which has axes aligned with the rotating frame and whose metric is related
to that of the rotating frame by Eq.~(\ref{rot_fields}).
Using the coordinate transformations (\ref{transf_law}) we can derive the
relation between all the fields in both frames, for example
\begin{eqnarray}
 u^0 & = & \bar{u}^0 \nonumber \\
 u^i & = & \bar{u}^i - (\vec{\Omega} \times \vec{r})^i \bar{u}^0 \nonumber \\
 u_i & = & \bar{u}_i \nonumber \\
 v^i & = & \bar{v}^i - (\vec{\Omega} \times \vec{r})^i,
\label{rot_matter}
\end{eqnarray}
where $v^i$ is the fluid three-velocity $v^i \equiv u^i/u^0$.  
At $t=\bar{t}=0$,  the
components of any spatial tensor are unchanged under this
transformation, for example
\begin{eqnarray} 
\gamma_{ij} = \bar{\gamma}_{ij},
\label{spatial_tensor}
\end{eqnarray}
and equivalently for $T_{ij}$ and $K_{ij}$.  The relation
(\ref{spatial_tensor}) implies $\gamma^{ij} = \bar{\gamma}^{ij}$,
since the inverse of a tensor is unique, so that we also find $K^{ij}
= \bar{K}^{ij}$. To complete our introduction of rotating frames in
general relativity, we refer the reader to Appendix \ref{newt_limit},
where we show that the Newtonian limit of the relativistic Euler
equation with a shift vector of the form of equation
(\ref{rot_fields}) reduces to the familiar Newtonian form of the Euler
equation in rotating frames.  In Appendix \ref{ADM_rotating}, we show
that the integrands used to evaluate $M$, $M_0$, and $J$ in Eqs.
(\ref{Mfin}), (\ref{Jfin}), and (\ref{M0fin}), respectively, remain
unchanged when expressed in terms of rotating frame variables.


\subsection{Boundary Conditions}
\label{gauge_BCs}

We always choose initial data which satisfy maximal slicing
(\ref{maximal_slicing}) and gauge choices which approximately maintain
this slicing.  Far from the source, equation (\ref{maximal_slicing})
becomes the Laplace equation, and its solution can be written as a sum
of multipole moment fields.  In the presence of matter, the source
will always have a nonzero monopole moment, so the asymptotic form of
the lapse is
\begin{equation}
\label{lapse_BC}
\alpha - 1 \propto r^{-1}.
\end{equation}

All of our spatial gauge choices resemble one another, so we will just
derive the shift boundary condition for the AMD shift (\ref{amd}),
which is the easiest.  The three components of Eq. (\ref{amd}) can be
decoupled by decomposing $\beta^i$ as in~\cite{s99}
\begin{equation}
\beta^i = \delta^{ji}\left[{7\over 8}P_i
  - {1\over 8}(\eta_{,i}+P_{k,i}x^k)\right],
\end{equation}
where $x^k$ are the Cartesian coordinates.  Eq.~(\ref{amd}) then becomes
\begin{eqnarray}
\label{AMD_P}
\nabla^2P_i &=& J_i \\
\label{AMD_eta}
\nabla^2\eta &=& - J_ix^i.
\end{eqnarray}
To lowest order, $J_i = \rho v^i$.  We will be studying systems with
azimuthal velocity fields, for which $v^z = 0$, and hence $\eta = P_z = 0$. 
The lowest nonvanishing moment of $J_i$, from the monopole piece of
$\rho$, is $l = 1$, $m = \pm 1$.  We can solve the Laplace equation
(outside the star) assuming asymptotic flatness to get
the boundary conditions $\beta^x \propto y r^{-3}$,
$\beta^y \propto x r^{-3}$, and, to this order, $\beta^z = 0$. 
A nonzero boundary condition
for $\beta^z$ must come from a higher-order term, but, since $\beta^z$
will be very small, our simulations are insensitive to it.  We use
\begin{equation}
\label{shift_BC}
\beta^x \propto y r^{-3},\qquad \beta^y \propto x r^{-3}, \qquad
\beta^z \propto x y z r^{-7}.
\end{equation}
The $\beta^z$ condition is obtained by ignoring $\eta$ and solving
(\ref{AMD_P}) subject to the lowest-order moment of the
$A_{ij}\alpha^{,j}$ term in $J_i$ (see Eq.~(\ref{amd_source}).)

Note that the coordinate-rotation component of a shift, $\vec{\Omega}
\times \vec{r}$, is a homogeneous solution of equation (\ref{amd}).
It was eliminated in (\ref{shift_BC}) by the assumption of asymptotic
flatness.  When working in a rotating frame, the shift does not obey an
asymptotic flatness condition.  Indeed, one can think of a
coordinate rotation as a boundary condition imposed on the
shift.  In such frames, the asymptotic form of the shift is
$\vec{\Omega}\times \vec{r}$ {\it plus} a piece which behaves like
(\ref{shift_BC}), and the boundary conditions must be set accordingly.


\section{Diagnostics}
\label{diag}

In order to gauge the accuracy of our simulations, we monitor the L2
norms of the violation in the constraint equations.  These are the
Hamiltonian constraint $\mathcal{H}$ ~(\ref{Hamiltonian_BSSN}), the
momentum constraint ${\mathcal{M}}^i$ (\ref{momentum_BSSN}), and the
Gamma constraint (\ref{gamma_BSSN}).  We normalize the Hamiltonian
and momentum constraint violation by their L2 norm by
\begin{eqnarray}
\label{Hnorm}
  N_{HC} &=& \left\|\left( \left(2\pi\psi^5\rho\right)^2 
  + \left(\tilde D^i\tilde D_i\psi\right)^2
  + \left({\psi\over 8}\tilde R\right)^2 \right.\right. \\
  & & + \left.\left. \left({\psi^5\over 8}\tilde A_{ij}\tilde A^{ij}\right)^2
     + \left({\psi^5\over 12}K^2\right)^2 \right)^{1/2}\right\|_2 \nonumber
\end{eqnarray}
and 
\begin{eqnarray}
\label{Mnorm}
 N_{MC} &=& 
 \left\|\left(\sum_{i=1}^3\left[(8\pi S^i)^2 
    + \left({2\over 3}\tilde D^iK\right)^2 \right.\right.\right. \\
  & & + \left.\left.\left.
    \left(\psi^{-6}\tilde D_j(\psi^6\tilde A^{ij})\right)^2
    \right]\right)^{1/2}\right\|_2.\nonumber
\end{eqnarray}
The two terms in the Gamma constraint (\ref{gamma_BSSN}) often vanish
individually, so that a similar normalization is not meaningful for
this constraint.

Related to the Hamiltonian and momentum constraints are mass and
angular momentum conservation. In Cartesian coordinates, the ADM mass
$M$ and the angular momentum $J^i$ are defined by the behavior of the
metric on a closed surface at asymptotically flat spatial infinity
\begin{eqnarray}
\label{Mdef}
M &=& {1\over 16\pi}\int_{r=\infty}\sqrt{\gamma}\gamma^{im}\gamma^{jn}
                                 (\gamma_{mn,j} - \gamma_{jn,m}) d^2S_i \\
\label{Jdef}
 J_i &=& {{1}\over{8\pi}}\varepsilon_{ij}{}^k \int_{r=\infty}
                         x^j K_k^m d^2S_m.
\end{eqnarray}
Since Eq.~(\ref{Jdef}) is computed at spatial infinity,
$\varepsilon_{ij}{}^k$ is the flat-space Levi-Civita tensor.  Using
Gauss' Law, we transform the surface integrals into volume integrals
\begin{eqnarray}
\label{Mfin}
M   &=& \int_V \Bigl(e^{5\phi}(\rho + {1\over 16\pi}\tilde A_{ij}\tilde A^{ij}
                             - {1\over 24\pi}K^2) \\
    & &  \quad - {1\over 16\pi}\tilde\Gamma^{ijk}\tilde\Gamma_{jik}
			     + {1-e^{\phi}\over 16\pi}\tilde R\Bigr) d^3x
			     \nonumber \\
\label{Jfin}
J_i &=& \varepsilon_{ij}{}^k\int_V \Bigl({1\over 8\pi}\tilde A^j_k
        + x^j S_k \\
\nonumber
    & & \quad +  {1\over 12\pi}x^j K_{,k}
        - {1\over 16\pi}x^j\tilde\gamma^{lm}{}_{,k}\tilde A_{lm}\Bigr)
        e^{6\phi} d^3x
\end{eqnarray}
(see, for example, Appendix A in \cite{yo02} for a derivation).  Note
that since $\varepsilon_{ij}{}^k$ is outside the integral, it is still
the flat-space Levi-Civita tensor.  Baryon conservation
$(\rho_0u^{\mu})_{;\mu} = 0$ implies that the rest mass
\begin{equation}
\label{M0fin}
M_0 = \int \rho_\star d^3x
\end{equation}
is also conserved.  Due to the finite differencing in our hydrodynamic
scheme, $M_0$ is conserved identically except for matter flow off the
computational grid.  We therefore monitor $M_0$ only as a diagnostic
of how much matter flows through the outer boundaries.

Eqs.~(\ref{Mfin}) and (\ref{Jfin}) are only valid in asymptotically
flat spatial hypersurfaces and thus are not suited for use in rotating
reference frames.  However, the problem can be sidestepped quite
easily by calculating the mass and angular momentum in the inertial
frame, as functions of the dynamical variables of the rotating
frame. In Appendix \ref{ADM_rotating} we show that the integrals for
these conserved quantities are exactly the same when expressed in
terms of the rotating frame fields.

Another useful quantity to monitor is the circulation.
According to the Kelvin-Helmholtz theorem \cite{c79}, the relativistic
circulation 
\begin{equation}
{\cal C}(c) = \oint_{c} h u_{\mu} \lambda^{\mu} d\sigma 
\end{equation}
is conserved in isentropic flow along an arbitrary closed curve $c$
when evaluated on hypersurfaces of constant proper time.  Here
$h=1+\varepsilon+P/\rho$ is the specific enthalpy, $\sigma$ is a
parameter which labels points on $c$, and $\lambda^{\mu}$ is the
tangent vector to the curve $c$.  Since ${\cal C}(c)$ is only
conserved for isentropic flow, checking conservation of circulation
along a few curves will measure the importance of numerical and
artificial viscosity on an evolution.  We do not monitor circulation
in this paper, although such a check has been implemented
elsewhere~\cite{ssbs01}.

Finally, we check for the existence of apparent horizons using the
apparent horizon finder described in~\cite{bcsst96}.


\section{Numerical Code Description}
\label{ABE}

All our algorithms have been implemented in a parallel,
distributed-memory environment using DAGH software~\cite{DAGH}
developed as part of the Binary Black Hole Grand Challenge Alliance.
When we need to solve elliptic equations (to construct initial data or
to impose elliptic gauge conditions), we use the computational toolkit
PETSc~\cite{petsc}.

\begin{figure}
\epsfxsize=2.8in
\begin{center}
\leavevmode \epsffile{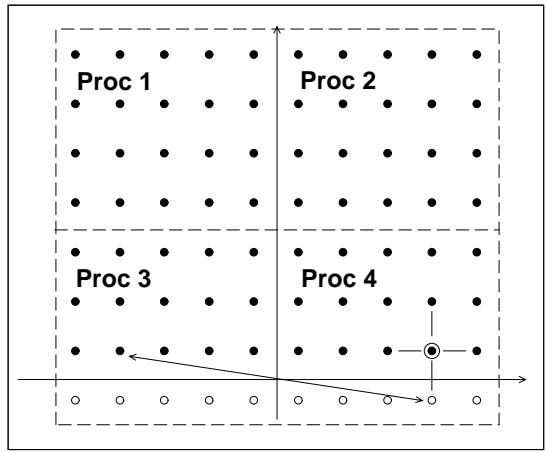}
\end{center}
\caption{This diagram shows how our code implements $\pi$-symmetry in 
distributed-memory computer clusters. The black circles correspond to grid 
points, and the bottom row corresponds to the boundary in the plane
orthogonal to the rotation axis. The white circles represent the 
ghostzones needed by our second order finite difference stencil. The arrow 
connects two points that are related in the presence of $\pi$-symmetry.}
\label{pi_symm}
\end{figure}

Due to our large number of variables, the memory needed by our code is
considerable (for example, a run with $64^3$ spatial zones may require
up to 2 Gbytes of
memory).  Thus, it is crucial that we exploit any symmetries present
in a given problem to minimize the number of gridpoints needed.  We
have implemented reflection symmetry across a coordinate plane
(equatorial symmetry) and reflection symmetry across three coordinate
planes (octant symmetry), which cut the size of our grids by factors
of two and eight.  Our code also allows us to enforce $\pi$-symmetry,
which assumes symmetry under a rotation of $\pi$ radians about a given
axis.  Unlike equatorial and octant symmetry, the implementation of
$\pi$-symmetry is not trivial on distributed-memory parallel systems.
This is because grid points needed to generate the proper boundary
conditions at a given location of the outer grid boundary will usually
be located in the memory of a different processor, as seen in the
diagram of Fig.~\ref{pi_symm} where the value of the field at the
white circle needed by point $P$ in processor $4$ must be provided by
a black circle on processor number $3$.  We fix this problem by
creating a two dimensional array for each field that stores the values
of the field on all the grid points outside the boundary (white
circles) needed to calculate the derivatives of the field at the grid
points at the boundaries (first row of black points).  Each processor
is responsible for updating the array values corresponding to grid
points within its domain by a $\pi$-radian rotation.  Updated values are
broadcast via MPI, and each processor has a copy of the complete
two-dimensional array from which to draw the corresponding boundary
values.


\section{Tests}
\label{tests}

\subsection{Vacuum Code Tests}

The algorithm for evolving the field equations was first tested in the
context of small amplitude gravitational waves~\cite{bs98b}.  With
harmonic slicing, the system could be accurately evolved for over 100
light crossing times without any sign of instability.  The results
also showed second-order convergence to the analytic solution when
resolution was increased.  In a forthcoming paper~\cite{yo02}, it will
be demonstrated that this same code can stably evolve isolated black
holes, with and without rotation, in Kerr-Schild coordinates.


\subsection{Hydro-without-Hydro}

Next, it was demonstrated that this field evolution scheme is stable
when predetermined matter sources are present~\cite{bhs99}.  This was done by
inserting the matter sources from known solutions of the Einstein
equations and then evolving the gravitational field equations.  Using
this ``hydro-without-hydro'' approach, \cite{bhs99} evolved the
Oppenheimer-Volkoff solution for static stars without encountering any
instability, and the Oppenheimer-Snyder solution for collapse of
homogeneous dust spheres well past horizon formation.  The same
hydro-without-hydro approach was later used to model the
quasi-equilibrium inspiral of binary neutron star systems
and calculate the complete late-inspiral gravitational wavetrain outside
the ISCO \cite{dbs01,dbssu02}.


\subsection{Shock Tube}
\label{shock_tube}

\begin{figure}
\epsfxsize=2.8in
\begin{center}
\leavevmode \epsffile{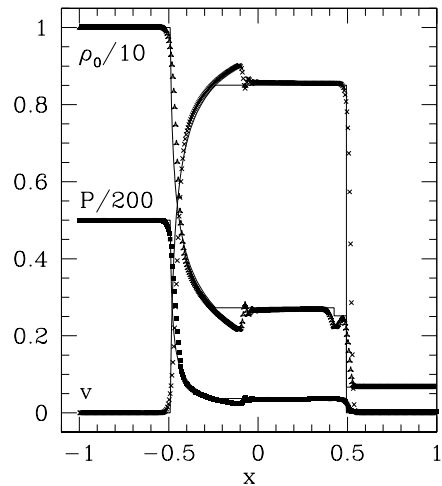}
\end{center}
\caption{ The one-dimensional relativistic Riemann shock tube test. 
  We plot the numerical rest density $\rho_0$ (triangles),
  pressure $P$ (squares), and velocity $v$ (crosses) at
  $t = 0.5$.  Solid curves show the analytic values.  This particular
  run used $C_{\rm Qvis} = 1$. }
\label{shock}
\end{figure}

Every hydrodynamic algorithm must demonstrate some ability to handle
shocks.  In Fig.~\ref{shock}, we compare the output of our code for a
simple one-dimensional shock tube problem with the exact result, which
is known analytically in special relativity~\cite{t86}.  In order to
compare with this result, the metric functions are held at their
Minkowski values throughout this test.  At $t = 0$, we set $v \equiv
v^x = 0$ everywhere.  For $x<0$ we set $\rho_0 = 15$, $P = 225$
initially, and for $x>0$ we set $\rho_0 = 1$, $P = 1$.  We output data
at $t = 0.5$.  In Figure 1, we use artificial viscosity parameters
$C_{\rm Qvis} = 1$, $C_{\rm Lvis} = 0$ (see Sec.~\ref{art_vis}) and a
grid with 400 points.  The shock is resolved quite well, and the only
disturbing feature of our results is the ``overshoot'' in variables at
the rarefaction wave.  Norman and Winkler~\cite{nw83} have shown that
these overshoots are present in the solution to the finite difference
equations of artificial viscosity schemes even in the limit of the
grid spacing going to zero.  This problem therefore represents a
fundamental limitation of artificial viscosity schemes, and points to
the need for more sophisticated high-resolution-shock-capturing
techniques when strong shocks are present (see, e.g., \cite{fmst00}).
However, for many of our astrophysical applications (e.g.~binary
inspiral) we anticipate at most very weak shocks, so that the use of
artificial viscosity schemes is adequate.

Our results are completely insensitive to $C_{\rm Qvis}$ when it is
within the range $0 - 0.1$.  We find the optimal behavior around
$C_{\rm Qvis} \approx 1$, at which point the effects of artificial
viscosity are small but noticeable.  For $C_{\rm Qvis}\approx 5$ or
greater, the viscosity is too large, and we are unable to evolve
accurately.  

Note that in the above example $\rho_{\star}>10^{-2}\rho_{\star \rm
max}$ everywhere, so $e_{\star}$-limiting (\ref{elimit}) is never
used.  More extreme shocks can be created by increasing the density
ratio $\rho_0(x>0)/\rho_0(x<0)$.  We find that we can treat shocks
reasonably accurately for ratios of up to about twenty.


\subsection{Oppenheimer-Snyder dust collapse}
\label{OS_coll}

\begin{figure}
\epsfxsize=2.8in
\begin{center}
\leavevmode \epsffile{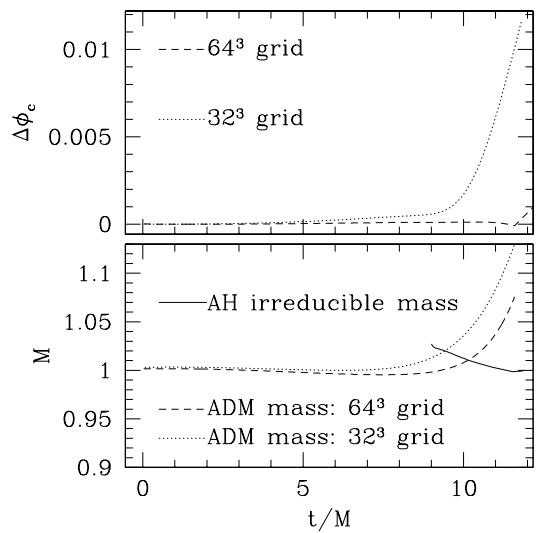}
\end{center}
\caption{ The evolution of the conformal exponent at the origin
  $\phi_c$ and the ADM mass $M$ during Oppenheimer-Snyder collapse. 
  The deviation of $\phi_c$ from its analytic value $\phi^{\rm anal}_c$ is
  measured by
  $\Delta\phi_c = (\phi_c - \phi^{\rm anal}_c)/(\phi_c + \phi^{\rm anal}_c)$. 
  This is plotted on the top panel.  On the bottom panel, we plot the ADM
  mass of the system and the irreducible mass of black hole, given by the
  area of the apparent horizon.  We compare runs at two different
  resolutions.}

\label{OS_phi}
\end{figure}

\begin{figure}
\epsfxsize=2.8in
\begin{center}
\leavevmode \epsffile{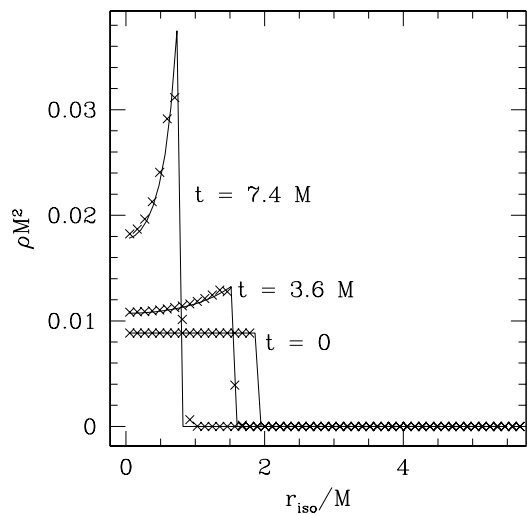}
\end{center}
\caption{ The density, defined by Equation~(\ref{rho_def})
  as a function of isotropic radius during Oppenheimer-Snyder
  collapse.  We compare our numerical results (crosses) with
  the analytic profiles. }
\label{OS_prof}
\end{figure}

As a second simulation which can be tested against exact results, we
model Oppenheimer-Snyder (OS) collapse of a homogeneous dust sphere to a
black hole~\cite{os39}.  The analytic solution for OS collapse can
be transformed into maximal slicing and isotropic
coordinates following~\cite{pst85}.  We use the analytic solution at
$t = 0$, when the matter is at rest, as initial data for all
variables.  We then evolve the gravitational and hydrodynamic fields
with our 3+1 code and compare the result with the exact solution.  At
each timestep, we determine the lapse by solving the maximal slicing
condition from the fields on our 3D grid.  For the shift we insert the
analytic values corresponding to isotropic coordinates.  We evolve on
a $32^3$ grid and a $64^3$ grid, utilizing octant symmetry to treat
only the upper octant.  Our outer boundaries are placed at $4M$ in
the isotropic coordinates of our grid.  The initial Schwarzschild
radius of the dust sphere is $3M$.

In Fig.~\ref{OS_phi}, we show the convergence of the central conformal
exponent $\phi_c$ to the exact value.  In Fig.~\ref{OS_prof}, we
compare the density profiles at several times for the $64^3$ grid to
their analytical values.  Throughout the evolution, we search for
apparent horizons.  At $t = 8.75 M$, we locate an apparent horizon
with irreducible mass $M_{AH}/M= 1.03$.  (See Fig.~\ref{OS_phi}.)
This mass remains constant to within 3\% until the end of the
simulation ($\approx 3 M$ later).  As is known analytically, all of
the mass falls inside the black hole.

This test is similar to the ``hydro-without-hydro'' Oppenheimer-Snyder
test performed on our code in~\cite{bhs99}, except that here the
matter fields and the lapse are determined numerically rather than set
to their analytic values.


\section{Single Stars}
\label{iso_stars}

In this Section we study isolated stars, both non-rotating and
rotating.  The initial data, constructed from the OV solution for
non-rotating equilibrium stars and with the code of \cite{cst92} for
equilibrium rotating stars, are summarized in Table~\ref{star_table}
(see also Fig.~\ref{sequences}.)  We use the same coordinates as used in
\cite{cst92} (except transformed from spherical to Cartesian). 
For spherical, nonrotating systems (OV stars), these are the
familiar isotropic coordinates.  In these coordinates, the 3-metric for
stationary spherical stars is conformally flat, and the event horizon
of a Schwarzschild black hole is located at $r = 0.5M$.
All stars are $n=1$, $\Gamma=2$ polytropes (see Eq.~(\ref{polytrope})),
and are dynamically evolved with the gamma-law equation of
state~(\ref{ideal_P}).  The nondimensional units throughout are set
by requiring $\kappa = G = c = 1$.

\newcommand{\stara}{A}
\newcommand{\starc}{B}
\newcommand{\stard}{C}
\newcommand{\stare}{D}
\newcommand{\starf}{E}
\newcommand{\starg}{F}

\begin{table}
\caption{Isolated Equilibrium Star Configurations ($\Gamma = 2$).}
\begin{tabular}{ccccccccc}
\hline \hline  
 Star & $M^a$ &
 $\left.\right.\ M_0{}^b\ \left.\right.$ 
 &$\left.\right.\ R_{eq}{}^c\ \left.\right.$ & $R_c{}^d$
 & $J/M^2{}^e$ & $T/|W|^f$ & $\Omega_c/\Omega_{eq}^g$ & $R_{pe}{}^h$
 \\ \hline  
\stara\ & 0.157 & 0.171 & 0.700 & 0.866 & 0.00 & 0.000 &      & 1.00 \\
\starc\ & 0.162 & 0.178 & 0.540 & 0.714 & 0.00 & 0.000 &      & 1.00 \\
\stard\ & 0.170 & 0.186 & 0.697 & 0.881 & 0.35 & 0.032 & 1.00 & 0.88 \\
\stare\ & 0.171 & 0.187 & 0.596 & 0.780 & 0.34 & 0.031 & 1.00 & 0.87 \\
\starf\ & 0.279 & 0.304 & 1.251 & 1.613 & 1.02 & 0.230 & 2.44 & 0.30 \\
\starg\ & 0.049 & 0.050 & 1.240 & 1.290 & 0.72 & 0.053 & 5.88 & 0.75 \\
\hline \hline  
\label{star_table}
\end{tabular}

\raggedright

${}^a$ ADM mass 

${}^b$ rest (baryonic) mass 

${}^c$ coordinate equatorial radius

${}^d$ areal radius at the equator

${}^e$ ratio of angular momentum to $M^2$

${}^f$ ratio of kinetic to gravitational potential energy

${}^g$ ratio of central to equatorial angular velocity

${}^h$ ratio of polar to equatorial coordinate radius

\end{table}

\begin{figure}
\epsfxsize=2.8in
\begin{center}
\leavevmode \epsffile{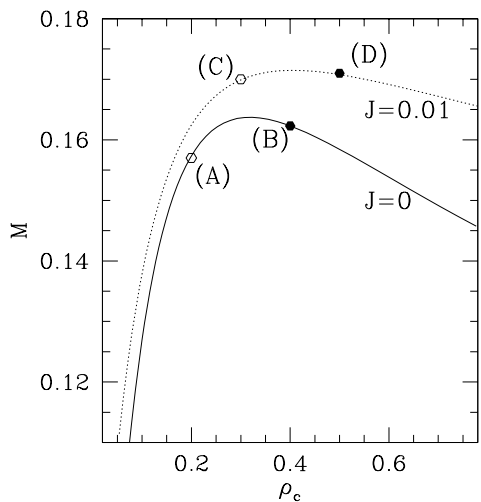}
\end{center}
\caption{ Stars \stara, \starc, \stard, and \stare\ and the
constant-$J$ sequences on which they lie.  Open circles represent
stable configurations, and closed circles denote unstable
configurations. }
\label{sequences}
\end{figure}


\subsection{Static Stars}
\label{TOV_stars}

The stability properties of non-rotating $\Gamma=2$ polytropes are
known analytically and can be used as a test of our code.  We use the
OV~\cite{ov39} solution describing equilibrium
polytropes in spherical symmetry as initial data, and evolve the
matter and fields dynamically.  An OV star is characterized by one
parameter, which can be taken to be the central rest density $\rho_c$.
(We will henceforth drop the subscript ``0'' on the rest density when
referring to central rest density.)  Along the sequence of increasing
$\rho_c$, the mass $M$ takes a maximum value $M_{\rm max}$ at a
critical central density $\rho_c^{\rm crit}$.  (See
Fig.~\ref{sequences}.)  Stars with $\rho_c < \rho_c^{\rm crit}$ are
dynamically stable, while stars with $\rho_c > \rho_c^{\rm crit}$ are
unstable and collapse to black holes on a dynamical timescale.  The
dynamical timescale is given by the free-fall time $\rho_c^{-1/2}$.
To verify that our code can distinguish stable and unstable
configurations we evolve two very similar models on either side of the
critical point at $\rho_c^{\rm crit}$.

\begin{figure}
\epsfxsize=2.8in
\begin{center}
\leavevmode \epsffile{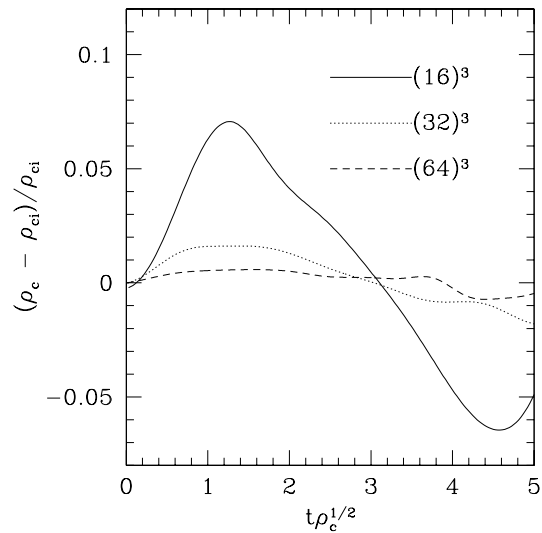}
\end{center}
\caption{Fractional change in the central rest density of star \stara\
when evolved on grids of three different resolutions.}
\label{converge}
\end{figure}

\begin{figure}
\epsfxsize=2.8in
\begin{center}
\leavevmode \epsffile{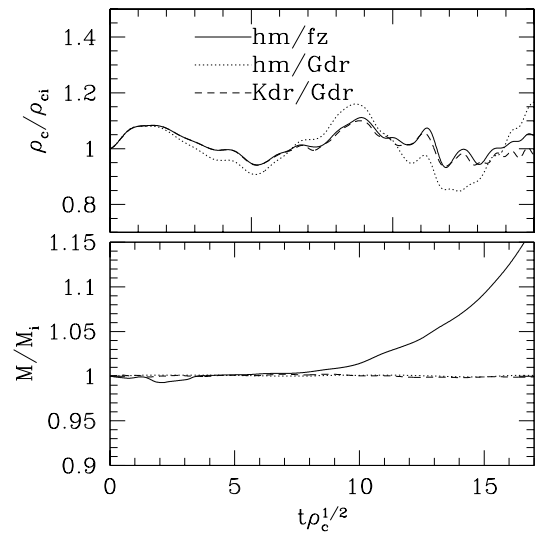}
\end{center}
\caption{ Star \stara\ evolved on a $32^3$ grid using various gauge
choices.  Here ``hm'' refers to harmonic lapse, ``Kdr'' to K-driver
lapse, ``fz'' to frozen shift, and ``Gdr'' to Gamma-driver shift. }
\label{TOV_gauge}
\end{figure}

In our units, $\rho_c^{\rm crit} = 0.32$ and $M_{\rm max} = 0.164$.
Star \stara\ has an initial central rest density $\rho_{ci} = 0.2$ and
is therefore stable. 
We set our outer boundaries at $x,y,z = 2$ and evolve this star with
three different resolutions $16^3$, $32^3$, and $64^3$, once again
utilizing octant symmetry.  In Fig.~\ref{converge}, we show the
central density evolution for the three resolutions using harmonic
slicing and the Gamma-driver shift.  We see that our code does
converge to the exact (stationary) solution.
There are three sources of the deviations from exact second-order
convergence (see also \cite{bhs99}).  First, there are components of
the error which scale with a higher power of the grid width
(e.g. $\Delta x^3$).  Second, there is the noise caused by
discontinuities at the surface of the star.  Finally, errors are
generated by imposing outer boundary conditions at finite distance.

In Fig.~\ref{TOV_gauge}, we evolve on a $32^3$ grid with several gauge
choices.  Already we see that the choice of gauge is important.  Even
in this static case, where the shift in the OV solution vanishes, it
is necessary to use a dynamic shift for long term stability.  With the
Gamma-driver, we evolve to $t\rho_c^{1/2} = 50$ ($t/M = 712$) and
never encounter an instability.

We have stably evolved star \stara\ on the $32^3$ grid for many 
fundamental radial oscillation periods, which have a period
of about $\tau_r =7\rho_c{}^{-1/2}$~\cite{Shibata:1999aa}.  However,
we find that
high-frequency, high-amplitude oscillations appear in $\rho_c$ after a
few periods and persist thereafter.  The onset of these oscillations
can be delayed and their amplitude diminished by increasing grid
resolution.  They can be removed altogether by making the hydrodynamic
algorithm more implicit, i.e. by increasing the weight on the new
timestep in the corrector step (See Section \ref{hydro_avd}).  This
adversely affects our ability to handle shocks, though.  The problem
may also be resolved by the use of a more sophisticated hydrodynamics
scheme (see, e.g., \cite{fmst00,fgimrssst02}).  For the less
relativistic stellar model used by \cite{fmst00,fgimrssst02} our code
produces non-physical high-frequency oscillations after about 6 radial
oscillation periods $\tau_r$.  For the applications discussed here,
these late-time problems are not relevant.

\begin{figure}
\epsfxsize=2.8in
\begin{center}
\leavevmode \epsffile{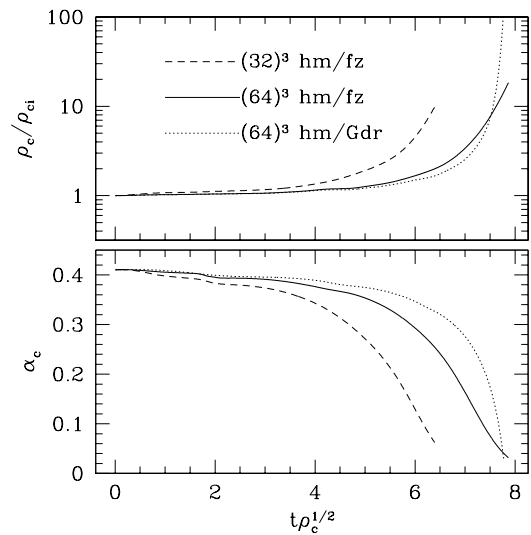}
\end{center}
\caption{ The collapse of star \starc\ seen with various gauge choices.  The
   abbreviations are the same as in Fig.~\ref{TOV_gauge}. }
\label{TOV_collapse}
\end{figure}

Star \starc\ has an initial central density of $\rho_{ci} = 0.4$ and
is dynamically unstable.  We evolve this star with harmonic lapse and
frozen shift, imposing outer boundaries at 1.2 $M$, on two different
grids ($32^3$ and $64^3$).  In Fig~\ref{TOV_collapse}, we plot the
central density and lapse as a function of time.  The collapse is
induced solely by the perturbations caused by putting the star on a
discrete grid.  Since these perturbations become smaller as grid
resolution is increased, it is not surprising that the star on the
lower-resolution grid collapses before the one on the
higher-resolution grid.  Since both collapse, it appears that $32^3$
zones are sufficient to distinguish stable from unstable stars.
Eventually, the star collapses to a point at which there are too few
grid points across the star's diameter for the evolution to remain
accurate.  We terminate our evolutions when the error in the ADM mass
exceeds 15\% of the original mass.  The $32^3$ grid turns out to be
too coarse for an apparent horizon to be located.  We do locate an
apparent horizon in the $64^3$ run shortly before the simulation is
terminated.  At this point the central lapse has collapsed to
$\alpha_c = 0.05$, and, as a measure of error, the ADM mass deviates
by 10\% from its initial value.  The horizon mass agrees well with the
ADM mass $M \approx M_{AH}$.

Also included in Fig.~\ref{TOV_collapse} is a $64^3$ simulation using
harmonic lapse and the Gamma-driver.  Similar behavior is seen in
these coordinates.  We will investigate the performance of various
gauge choices in more detail in the following Section.

\subsection{Uniformly Rotating Stars}
\label{uniform}

\subsubsection{Inertial Frame}
\label{rns_inert}

\begin{figure}
\epsfxsize=2.8in
\begin{center}
\leavevmode \epsffile{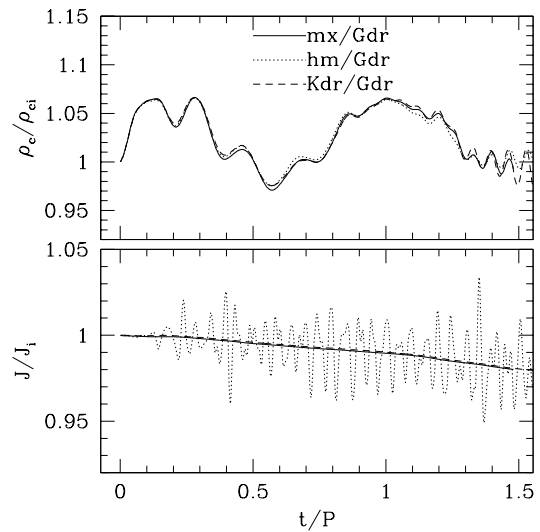}
\end{center}
\caption{ Star \stard\ evolved on a $64\times 32^2$ grid with Gamma-driver
  shift condition and various lapse choices. }
\label{lapse}
\end{figure}

\begin{figure}
\epsfxsize=2.8in
\begin{center}
\leavevmode \epsffile{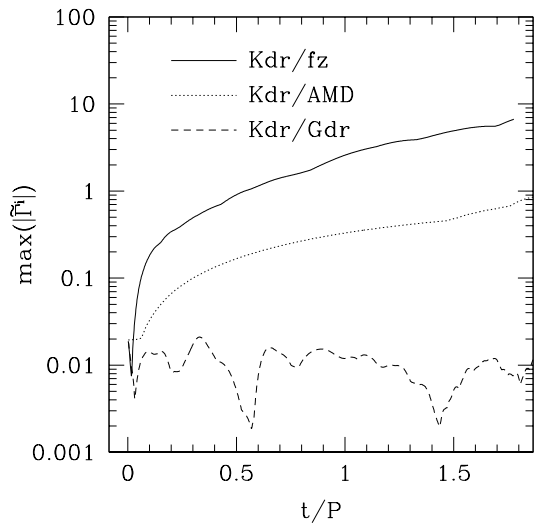}
\end{center}
\caption{ Star \stard\ evolved on a $64\times 32^2$ grid with K-driver
  lapse and various shift choices.  We plot the maximum value of the
  absolute value of $\tilde\Gamma^x$ on the grid to show the
  dependence of $\tilde\Gamma^i$ on spatial gauge. }
\label{shift}
\end{figure}

Simulations of systems with non-zero angular momentum are very
sensitive to the choice of coordinates, which makes them very good
test cases for comparing the numerical behavior of different gauges
and slicings.  Most of these effects can be seen when we evolve
uniformly rotating stars.

We consider two uniformly rotating stars, stars \stard\ and \stare, on
one constant angular momentum sequence $J = 0.01$ (see
Fig.~\ref{sequences}).  The $J = 0.01$ sequence has a turning-point at
$\rho_c^{\rm crit} = 0.4$, $M_{\rm max} = 0.172$.  For a sequence of
uniformly rotating stars, this turning point marks the onset of
secular, not dynamical, radial instability~\cite{fis88}.  It is possible for
a star on the secularly unstable branch to be stabilized temporarily
if the star begins to rotate differentially, so that no instability
will develop on a dynamical timescale.  However, prior numerical
studies~\cite{sbs00a} have found the point of onset of dynamical
instability to be very close to the point of onset of secular instability,
which we confirm with our simulations here.

We again pick two similar stars on either side of the onset of secular
instability: star \stard\ with $\rho_{ci} = 0.3$ on the stable branch
and Star \stare\ with $\rho_{ci} = 0.5$ on the unstable branch.  We
dynamically evolve these two stars with different choices for the
slicing and gauge.  All simulations are performed on $64\times 32^2$
grids, utilizing equatorial and $\pi$-symmetry to evolve only half of a
hemisphere.  The outer boundaries are placed at $[-1.5,1.5]\times [0,1.5]^2$. 
There are now two relevant timescales -- the free-fall time $\tau_{ff} \sim
\rho_c^{-1/2}$ and the orbital period $P$ -- and a reliable code must
be able to stably evolve stable rotating stars for several of both
timescales.

Results for star \stard\, with $\tau_{ff} = 1.83$ and $P = 26.38$ in
our units, are plotted in Figures~\ref{lapse}
and~\ref{shift}.  In Fig.~\ref{lapse} we compare the evolution for
maximal slicing (\ref{maximal_slicing}), harmonic slicing
(\ref{harmonic}) and the K-driver (\ref{K_driver}), all with the
Gamma-driver shift condition (\ref{gamma_driver}).  We find that there
is little sensitivity to the lapse choice except for small
oscillations in $J$ which are only present with harmonic slicing.

In Fig.~\ref{shift} where we compare the frozen shift condition
(\ref{frozen}), the AMD shift (\ref{amd}) and the Gamma driver
(\ref{gamma_driver}), all evolved with the K-driver (\ref{K_driver})
for the lapse.  This comparison demonstrates the great importance of
choosing an appropriate shift condition for controlling
$\tilde\Gamma^i$.  AMD is dramatically better than frozen shift in
this regard, and the Gamma driver is dramatically better than AMD.
The behavior with AMD shift does not change significantly when the
criteria for convergence of (\ref{amd}) is made stricter.  Note that
the modification (\ref{mamd}) to AMD and the Gamma-driver is not
activated for this application.

\begin{figure}
\epsfxsize=2.8in
\begin{center}
\leavevmode \epsffile{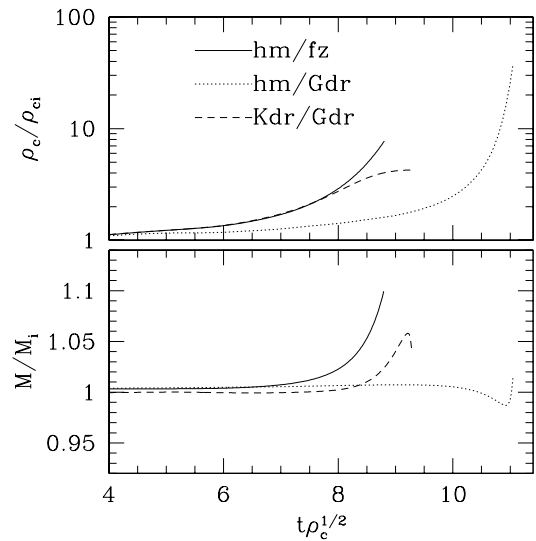}
\end{center}
\caption{ The evolution of the central rest density and the ADM mass as
  star \stare\ collapses. }
\label{rcoll1}
\end{figure}

\begin{figure}
\epsfxsize=2.8in
\begin{center}
\leavevmode \epsffile{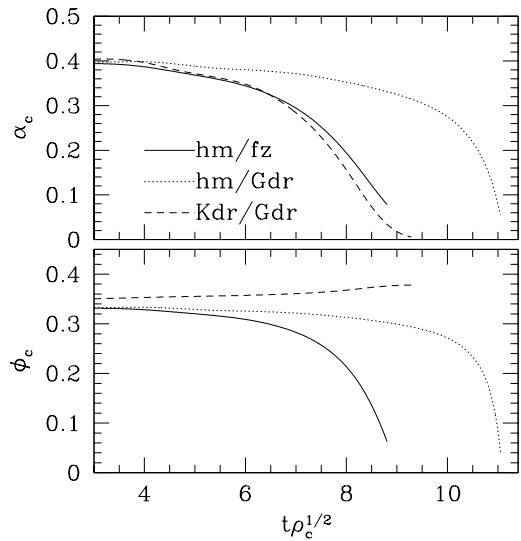}
\end{center}
\caption{ The evolution of the lapse and conformal exponent at the origin
  as star \stare\ collapses. }
\label{rcoll2}
\end{figure}

Figures~\ref{rcoll1} and~\ref{rcoll2} show the behavior of the
radially unstable star \stare\ under different coordinate choices. 
Once again, perturbations are induced solely by the finite difference
error of the grid.  We terminate simulations when mass conservation is
violated by 10\% or the code crashes.  The singularity avoidance property
of the K-driver, which approximates maximal slicing, is manifest: with
the lapse
collapsing to very small values, the proper time between time slices
at the star's center becomes very small, which effectively ``freezes''
all quantities there.  With harmonic slicing, $\alpha$ decreases more
slowly, and we are able to reach higher central densities, corresponding
to later proper times, before the code crashes.  Given their
qualitatively different behavior, it is difficult to compare meaningfully
the different lapse choices for this scenario.  If one wants
to see the central region reach the farthest stage of collapse before
violation of mass conservation becomes unacceptable, harmonic lapse
and Gamma-driver shift seem to be the optimal combination.  One
possible reason for this is the behavior of $\phi_c$, the conformal
exponent at the stellar center.  For the gauge choices which are best
suited to probing the central region, $\phi_c$ decreases
significantly from its initial value.  Inverting Shibata's reasoning
for modifying the AMD gauge, we infer that this corresponds to
choosing a gauge with infalling coordinates. This effectively
increases the grid coverage of the collapsing star, resulting in a
more accurate evolution.

We are only able to locate an apparent horizon in the harmonic
lapse/Gamma driver simulation, and only
in the last few timesteps, at which point $\alpha_c = 0.05$,
 $M_{AH}/M = 0.58$, and the
error in ADM mass is about 2\%.  It seems that $64\times 32^2$ zones are
barely sufficient resolution for locating horizons for rotating
stars reliably.


\subsubsection{Rotating Frame}
\label{rns_rotating}

\begin{figure}
\epsfxsize=2.8in
\begin{center}
\leavevmode \epsffile{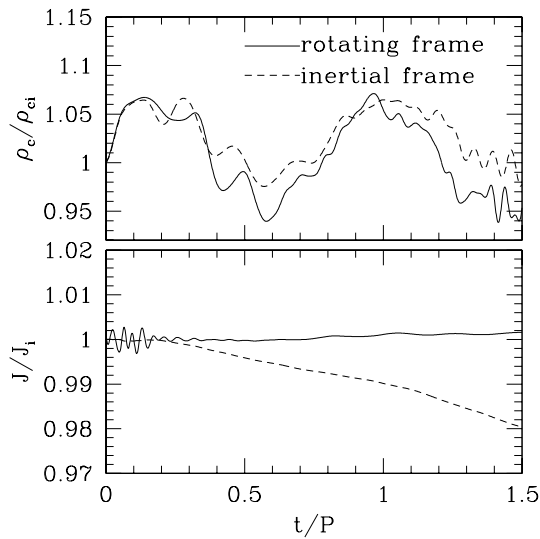}
\end{center}
\caption{ The central density and angular momentum of star \stard\ evolved
  on a $64\times 32^2$ grid with K-driver lapse and Gamma-driver shift
  in the inertial and in the rotating frame.  We see that $J$ is conserved
  much better in the rotating frame. }
\label{rns_IvsR}
\end{figure}

We compare results for uniformly rotating star \stard\ in the inertial
frame to results
in the corotating frame in Figure \ref{rns_IvsR}.  In the corotating
frame, $v^i = 0$ at $t=0$.  The light cylinder, where points of fixed
coordinate label are moving on null paths, is at $r_{\rm cyl} = 4.2$,
well outside our outer boundaries at $x,y,z = 1.5$.  All coordinate
observers are therefore timelike everywhere on our grid.  We see a
dramatic improvement in angular momentum conservation in the
corotating frame.  This indicates that the loss of $J$ in
the inertial frame is caused by error in the advection of fluid
quantities along $v^i$.  Mass conservation is also better in the
rotating frame, but not dramatically.  Other quantities show
qualitatively the same behavior in both frames.  We have redone the
evolution of collapsing star \stare\ with harmonic lapse and Gamma-driver
shift in the corotating frame, and our results were almost identical to
those in the inertial frame.


\subsection{Differentially Rotating Stars}
\label{differential}

\begin{figure}
\epsfxsize=2.8in
\begin{center}
\leavevmode \epsffile{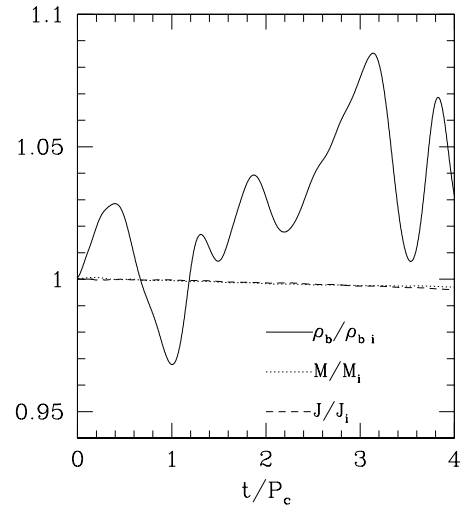}
\end{center}
\caption{ Star \starf\ evolved for 4 central periods on a $64\times
32^2$ grid. The $M$ and $J$ curves overlap. }
\label{Estarf}
\end{figure}

We now test the ability of our code to handle differential rotation.
Differential rotation in neutron stars is relevant in several
important astrophysical phenomena.  Simulations in both Newtonian
hydrodynamics~\cite{rs99} and full general relativity
\cite{Shibata:1999wm,Shibata:2002jb} indicate that binary neutron star
coalescence may well lead, at least temporarily, to a differentially
rotating remnant, which can support significantly more rest mass than
uniformly rotating stars~\cite{bss00}.  Core collapse in a supernova
may also result in a differentially rotating neutron star.

\begin{figure}
\epsfxsize=2.8in
\begin{center}
\leavevmode \epsffile{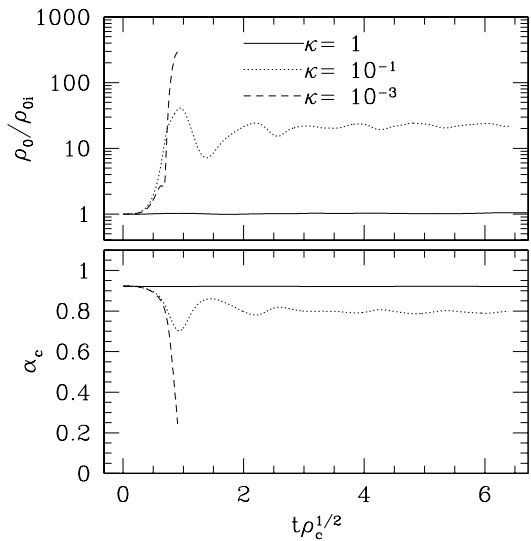}
\end{center}
\caption{ The evolution of star \starg\ with 0\%, 90\%, and 99.9\% of its
  pressure removed, respectively.  When no pressure is removed, the
  star is stationary.  When 90\% is removed, we evolve until the new
  equilibrium is reached.  When 99.9\% is removed, the star collapses
  from an initial radius of $26M$ to a radius of $\sim 5M$, at which
  point the simulation becomes inaccurate, and we terminate it. }
\label{Pdep}
\end{figure}

We construct axisymmetric equilibrium initial data, again following
\cite{cst92}, with $z$ chosen as the axis of symmetry.  For the rotation
profile, we choose
\begin{equation}
\label{diff_rot}
u^tu_{\phi} = R_{\rm eq}^2 A^2(\Omega_c - \Omega)
\end{equation}
where $\Omega$ is the angular velocity of the fluid, $\Omega_c$ is the
value of $\Omega$ on the rotation axis, $R_{\rm eq}$ is the equatorial
coordinate radius, and $A$ is a parameter that measures, in units of
$R_{\rm eq}$, the scale over which $\Omega$ changes.  In the Newtonian limit
this profile reduces to
\begin{equation}
\Omega = {A^2 \Omega_c\over (x^2 + y^2)/R_{\rm eq}^2 + A^2}.
\end{equation}
For $A \rightarrow \infty$ one recovers uniform rotation.

In Fig.~\ref{Estarf} we present results for star \starf\ with
$\rho_{\rm max} = 0.07$, $A^{-1} = 1$, $R_{\rm eq}/M = 4.48$, $T/|W| =
0.23$, and $J/M^2 = 1.02$.  This star's rest mass of $M_0 = 0.304$
exceeds the maximum allowed rest-mass of non-rotating $\Gamma = 2$
polytropes by 70  percent.  We evolve
this star on a $64\times 32^2$ grid, using $\pi$-symmetry, with outer
boundaries at $[-2,2]\times [0,2]^2$.  The same star was evolved
dynamically by~\cite{bss00}, and we confirm their finding that the
star is stable over several central rotation periods.

We found that simulations of differentially rotating stars are very
sensitive tests of hydrodynamic advection schemes.  In particular, when
we used time-averaging instead of Crank-Nicholson to treat the
advection terms (see Appendix \ref{advection_appendix}), we found that
the angular momentum is conserved very poorly.  The decrease in
$J$ also causes the central density to rise, and the numerical model
to drift further and further from the true solution.  This suggests
that for differential rotation, the ability to successfully conserve
angular momentum depends strongly on the finite difference algorithm
used for the hydrodynamics.

In Fig.~\ref{Pdep} we show results for star \starg, with $\rho_{\rm
max} = 0.0174$, $A^{-1} = 3$, $R_{\rm eq}/M = 26.3$, $T/|W| = 0.0528$,
and $J/M^2 = 0.715$.  This model is identical to star I in Table II of
\cite{Shibata00}, where this star was evolved in axisymmetry.  Star
\starg\ is radially stable, but, as in \cite{Shibata00}, we make the
situation dynamic by depleting pressure from the star by artificially
reducing the polytropic constant $\kappa$ (which requires us to re-solve
the Hamiltonian and momentum constraints).  Removing
pressure support causes the star to implode.  For small depletion
factors, this collapse is halted and the star bounces and finds a new, more
compact, stable equilibrium configuration.  When $\kappa$ is reduced
to a low enough value, the star will collapse to a black hole.  Thus,
there is a critical polytropic constant $\kappa_{\rm crit}$ separating
these two outcomes.  In \cite{Shibata00}, this critical value was
found to be $\kappa_{\rm crit} \approx 0.04 $.

\begin{figure}
\epsfxsize=2.8in
\begin{center}
\leavevmode \epsffile{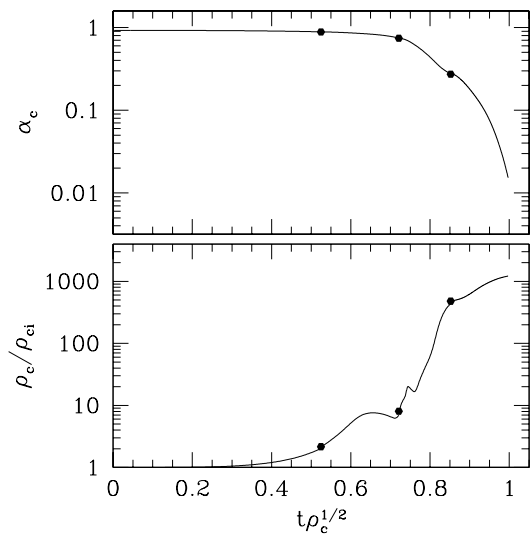}
\end{center}
\caption{ The evolution of star \starg\ with 99.9\% of its pressure removed. 
  This time, we evolve on a $100^2\times 50$ grid.  The points mark times
  when the resolution was doubled. }
\label{Collhr1}
\end{figure}

We evolve star \starg\ on a grid of $64^2\times 32$ zones, with
the outer boundaries located at 2, or equivalently 40.8 $M$. 
In this simulation we use only equatorial
symmetry so that non-$\pi$-symmetric perturbations can
grow.  We evolve three different cases; one without
pressure depletion with $\kappa = 1$, a supercritical case with
$\kappa = 0.1 > \kappa_{\rm crit}$, and a subcritical case with
$\kappa = 0.001 < \kappa_{\rm crit}$.  Both the second and third case
present unique challenges.  In the second case, the collapse is halted
by a strong shock which must be handled accurately.  In the third
case, we must follow the collapse from a radius of $26.3 M$ to a
radius of $\approx M$.  Our results are consistent with those of
\cite{Shibata00}, even though our 3D simulations have a much poorer
resolution than the axisymmetric simulations of \cite{Shibata00}.  In
particular, our resolution is insufficient to follow the final stages
of the $\kappa = 0.001$ collapse and prove that a black hole is
formed.

In order to overcome this problem, we redo the $\kappa = 0.001$ collapse
on a $100^2\times 50$ grid.  This grid is still too sparse to resolve
a black hole with radius of approximately $1M$ if the outer boundaries
are imposed at 40.8 $M$.  In order to resolve the black hole, we
rezone our grid several times during the implosion, halving the boundaries
and halving
the grid spacing, so that the total number of gridpoints remains
constant (compare~\cite{ss02}.)  We present results for a simulation
that was carried out on four different grids with outer boundaries at
2, 1 0.5 and 0.25.  We use the K-driver and the Gamma-driver without
the modification~(\ref{mamd}).  With the modification, the functions
$\tilde\Gamma^i$ grow very rapidly and cause the code to crash well
before the radius reaches $M$.  Turning off the modification means
that $\phi_c$ will grow, and the coordinates will blow outward.  We
count on the grid rezoning to counter this effect.  Also, we switch to
frozen shift on the last and finest grid of the evolution.  The
Gamma-driver does not perform well on this segment, perhaps because
the grid boundaries have been moved in to a point where the shift does
not have its asymptotic form (\ref{shift_BC}).

\begin{figure*}
\begin{center}
\epsfxsize=2.0in
\leavevmode
\epsffile{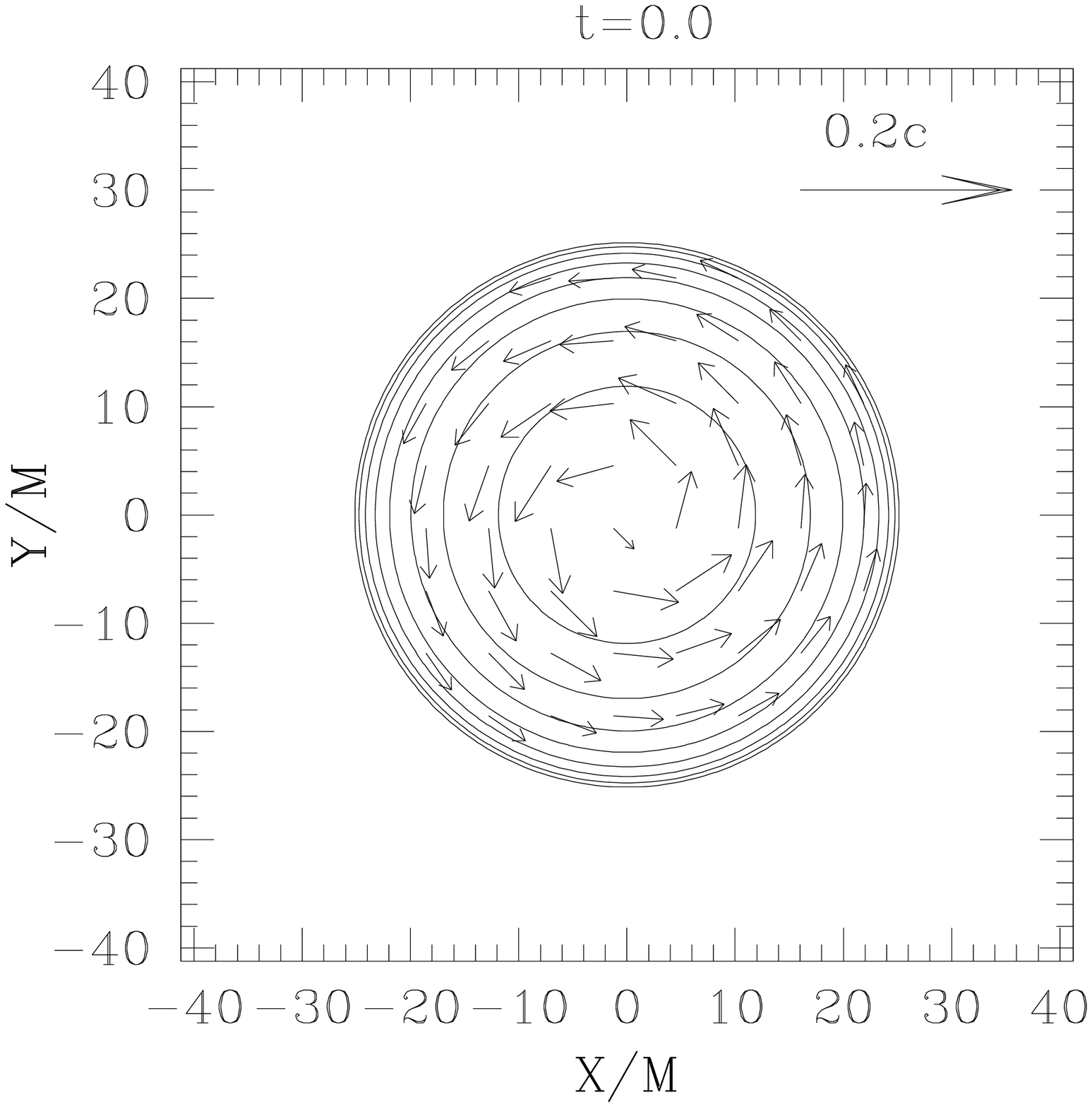}
\epsfxsize=2.0in
\leavevmode
\epsffile{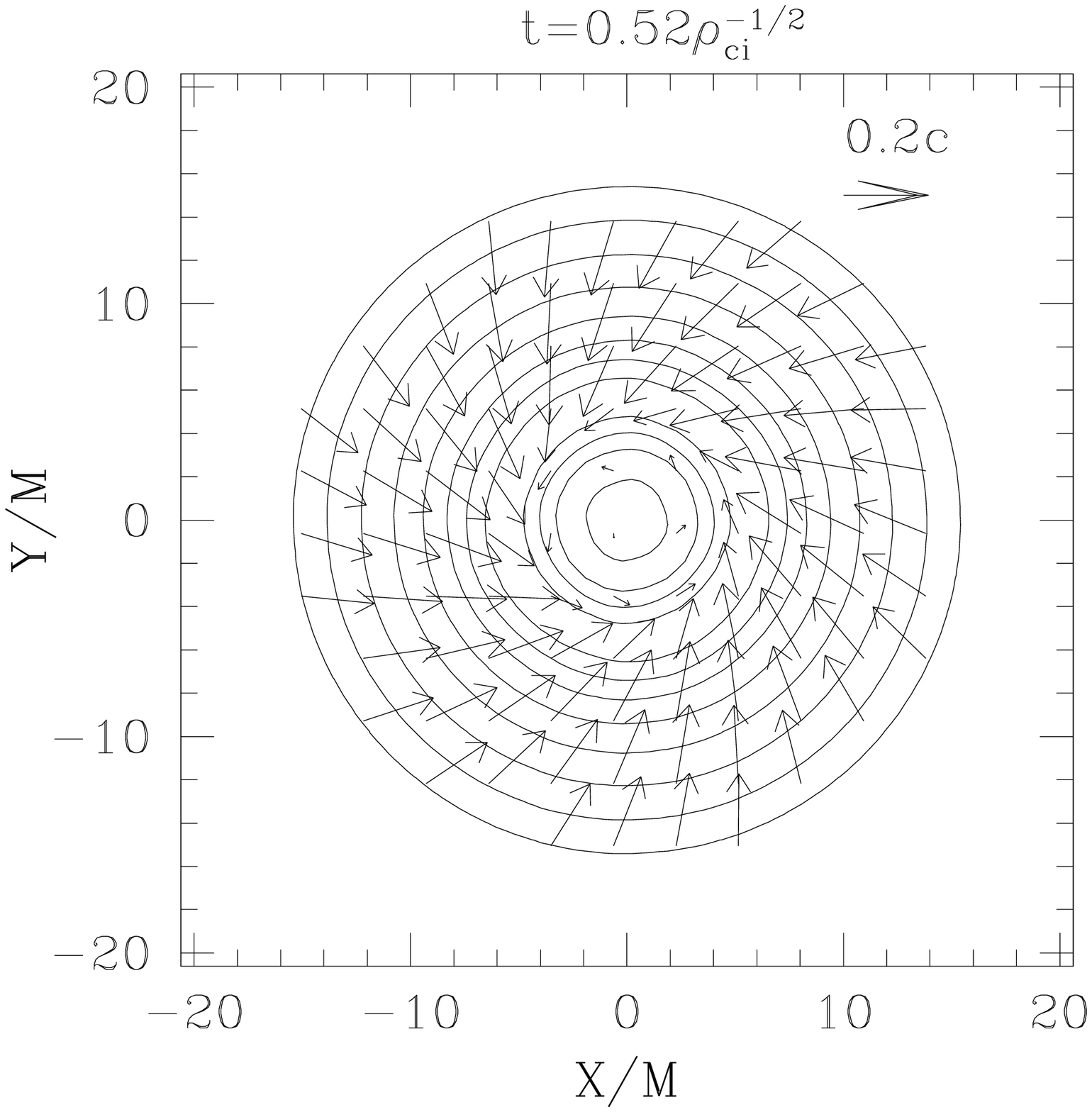}\\
\epsfxsize=2.0in
\leavevmode
\epsffile{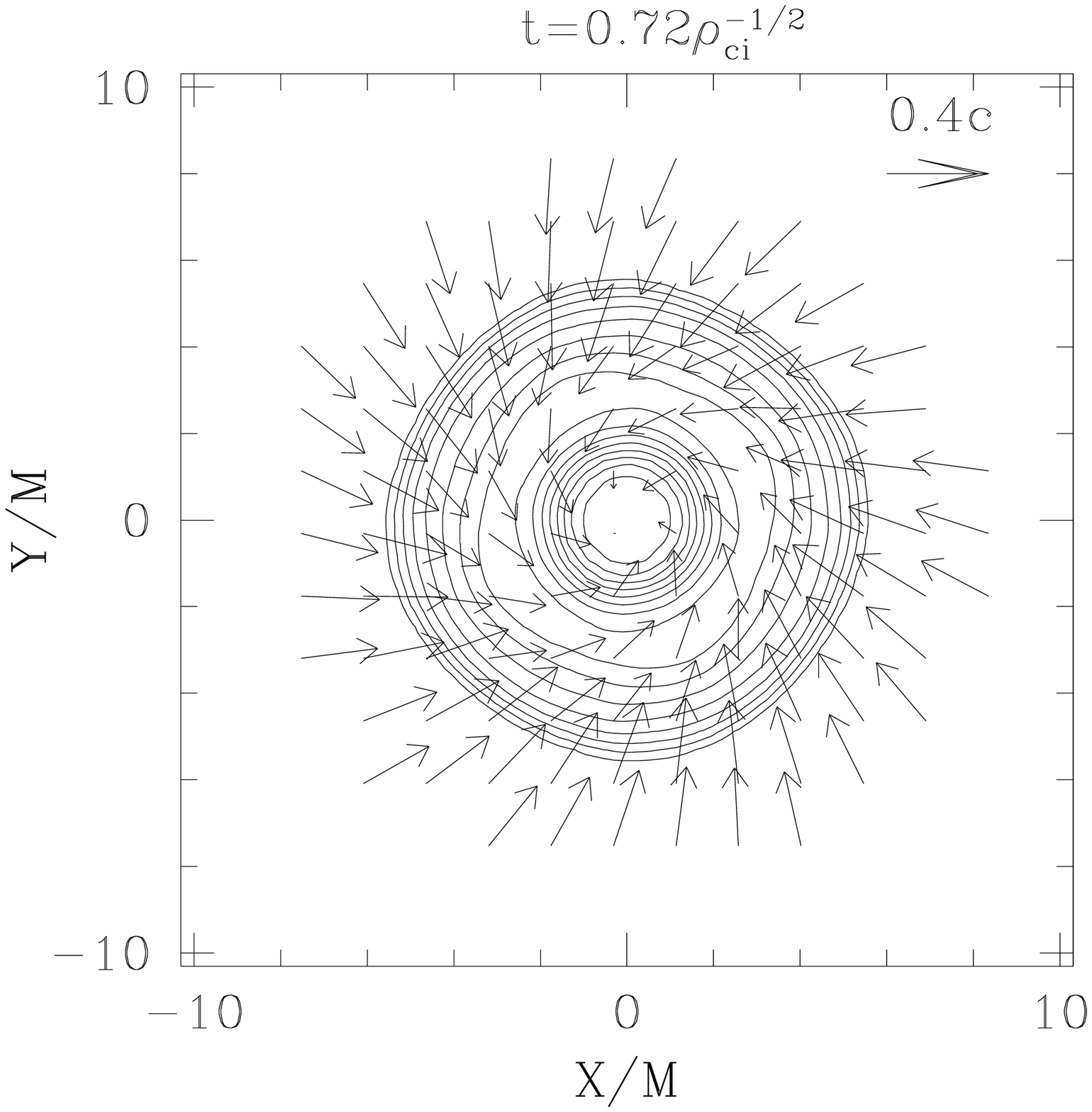}
\epsfxsize=2.0in
\leavevmode
\epsffile{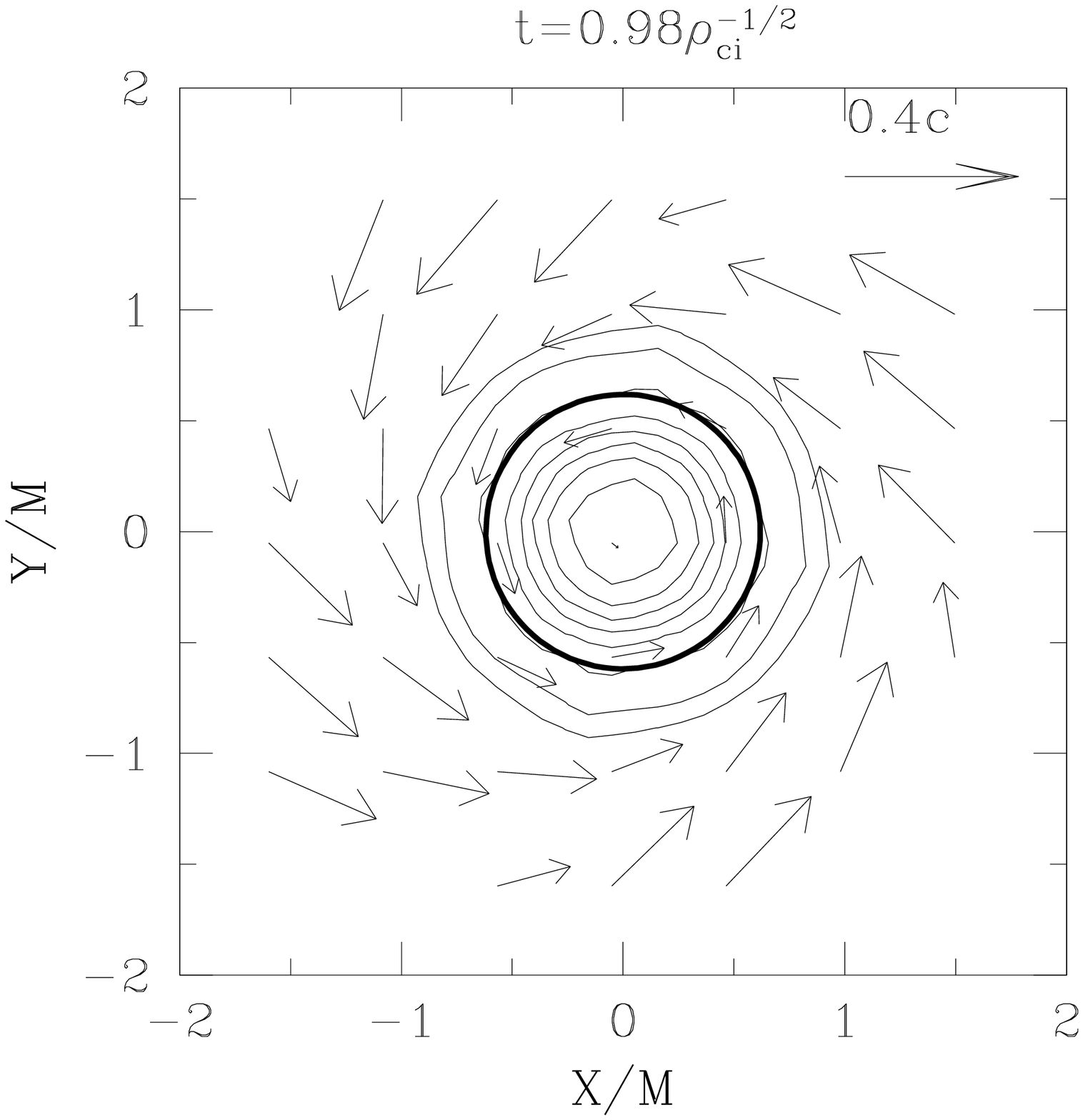}
\caption{ Snapshots of the rest density contour lines for $\rho_0$ and the
velocity field $(v^x,v^y)$ in the equatorial plane for a simulation of
the collapse of star \starg\ with $\kappa=0.001$.  The contour lines are
drawn for $\rho_0=10^{-(0.2~j+0.1)} \rho_{0}^{Max} $, where $\rho_{0}^{Max}$
denotes the instantaneous maximum value of $\rho_0$ for $j=0,1,..,7$.
Vectors indicate the local velocity field, $v^i$.  The thick circle on the
last frame marks the apparent horizon. }
\label{coll_contour}
\end{center}
\end{figure*}

The results of this simulation are shown in Figs.~\ref{Collhr1}
and~\ref{coll_contour}.  $M$ and $J$ remain within 10\% of their
initial values throughout (we terminate the calculation when this
ceases to be true.)  In our coordinates, the equatorial radius
decreases from 1.24
(25.3$M$) to 0.04 (0.8$M$).  Since $\phi_c$ is growing, part of this
decrease in radius is a coordinate effect.  The coordinate-independent
circumferential radius at the equator (computed from $g_{\phi\phi}$)
decreases from 27$M$ to 1.7$M$.  At $t\rho_c^{1/2} = 0.98$, we locate
an apparent horizon with surface area $\mathcal{A} =$ 0.0804
corresponding to an ``irreducible'' mass of
$(\mathcal{A}/16\pi^2)^{1/2} = 0.8M$.  Is this
a reasonable value?  The existence of rotation and
of mass outside of the black hole mean that we can no longer expect
the irreducible mass of the hole to be equal to the ADM mass of the
entire system.  The area of the event horizon of a Kerr black hole
with this system's total $M$ and $J$ would be $\mathcal{A} =$ 0.109.  By
breaking up the rest mass integral into pieces inside and outside the
horizon, we find that 82\% of the baryonic mass is inside the apparent
horizon.  If we assume that the values of $M$ and $J/M$ for the black
hole are 82\% of those of the total system, we arrive at the very
crude estimate $\mathcal{A} =$ 0.0732, which is within 10\% of the
value determined from the apparent horizon.  We terminate our
simulation 2.5$M$ after the horizon is located, during which time its
area does not change appreciably.

Our agreement with~\cite{Shibata00} indicates that nonaxisymmetric
perturbations are not important in the collapse of this star.  We
confirm this in Fig.~\ref{coll_contour}.  As one can see, the density
profiles remain axisymmetric throughout.


\section{Binary Systems}
\label{Bin}

\begin{figure*}
\begin{center}
\epsfxsize=2.6in
\leavevmode
\epsffile{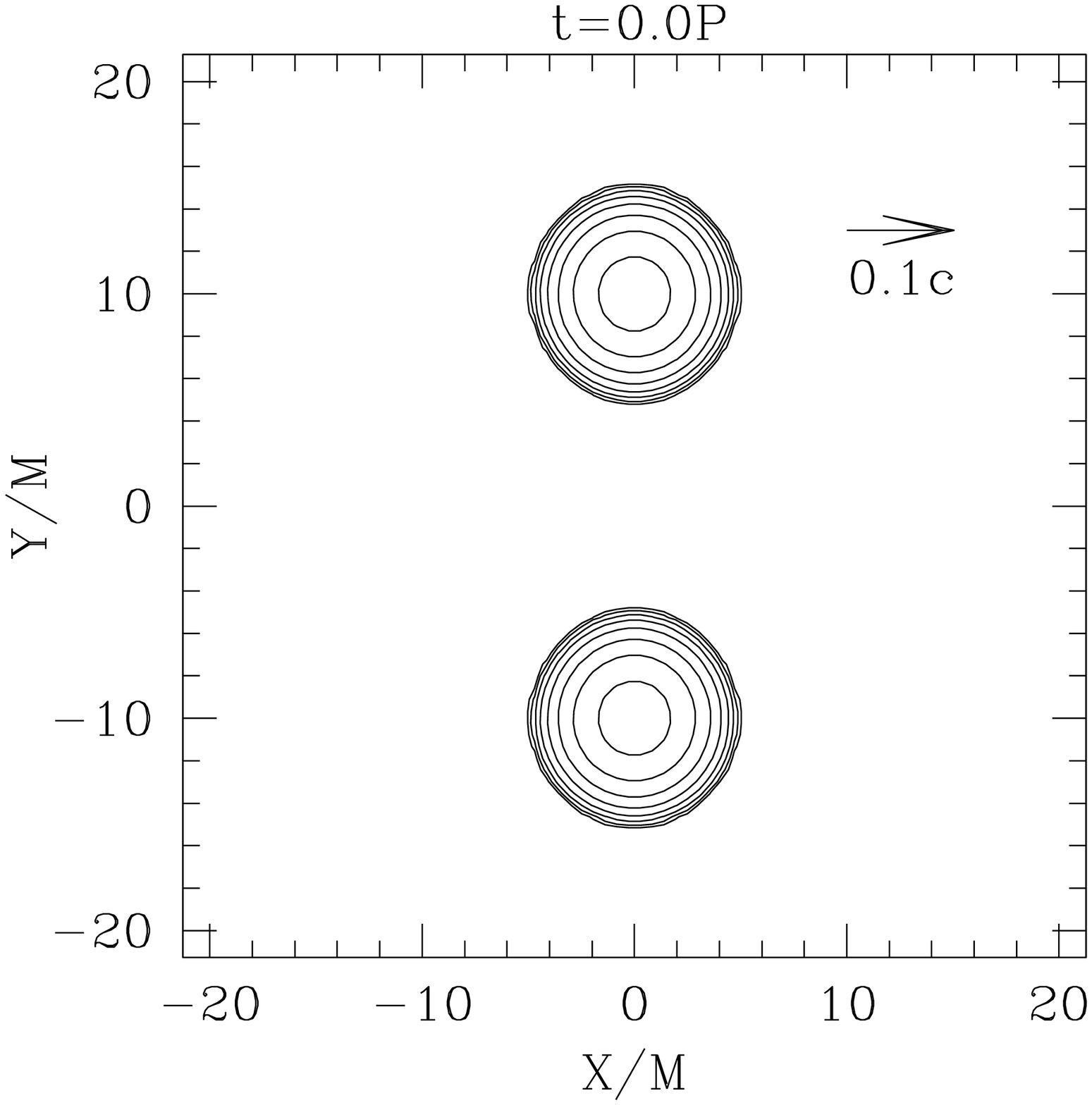}
\epsfxsize=2.6in
\leavevmode
\epsffile{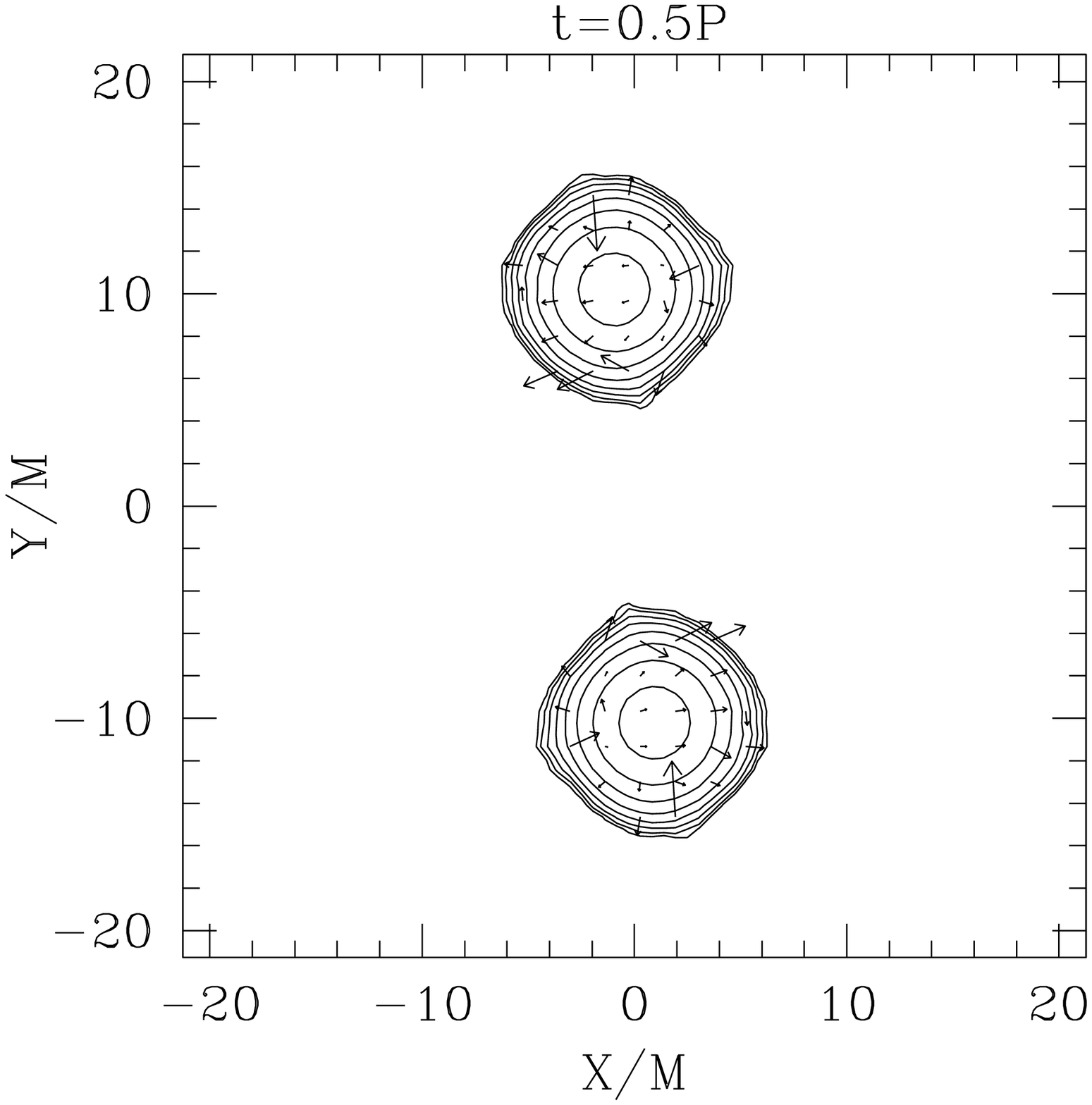}\\
\epsfxsize=2.6in
\leavevmode
\epsffile{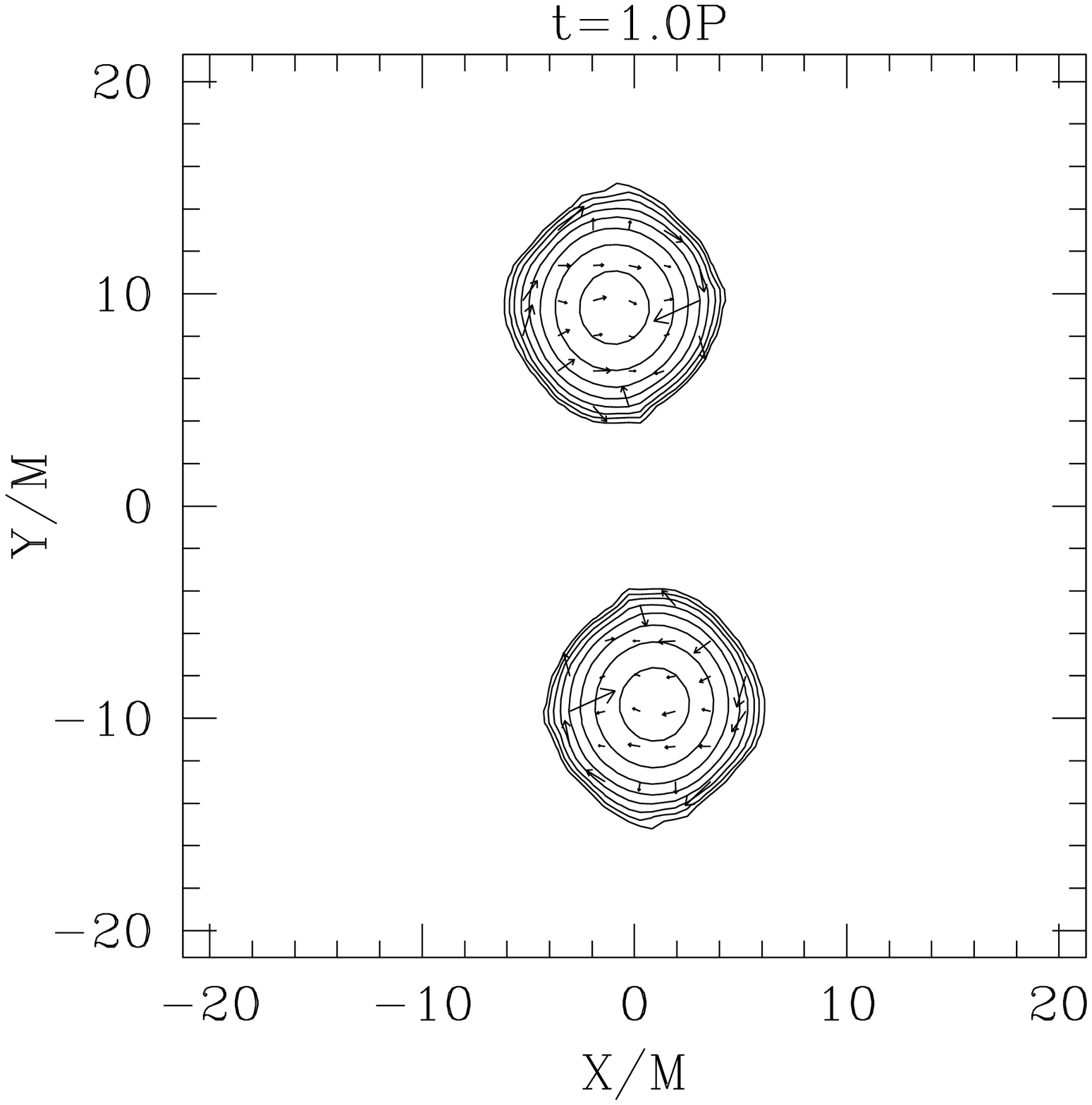}
\epsfxsize=2.6in
\leavevmode
\epsffile{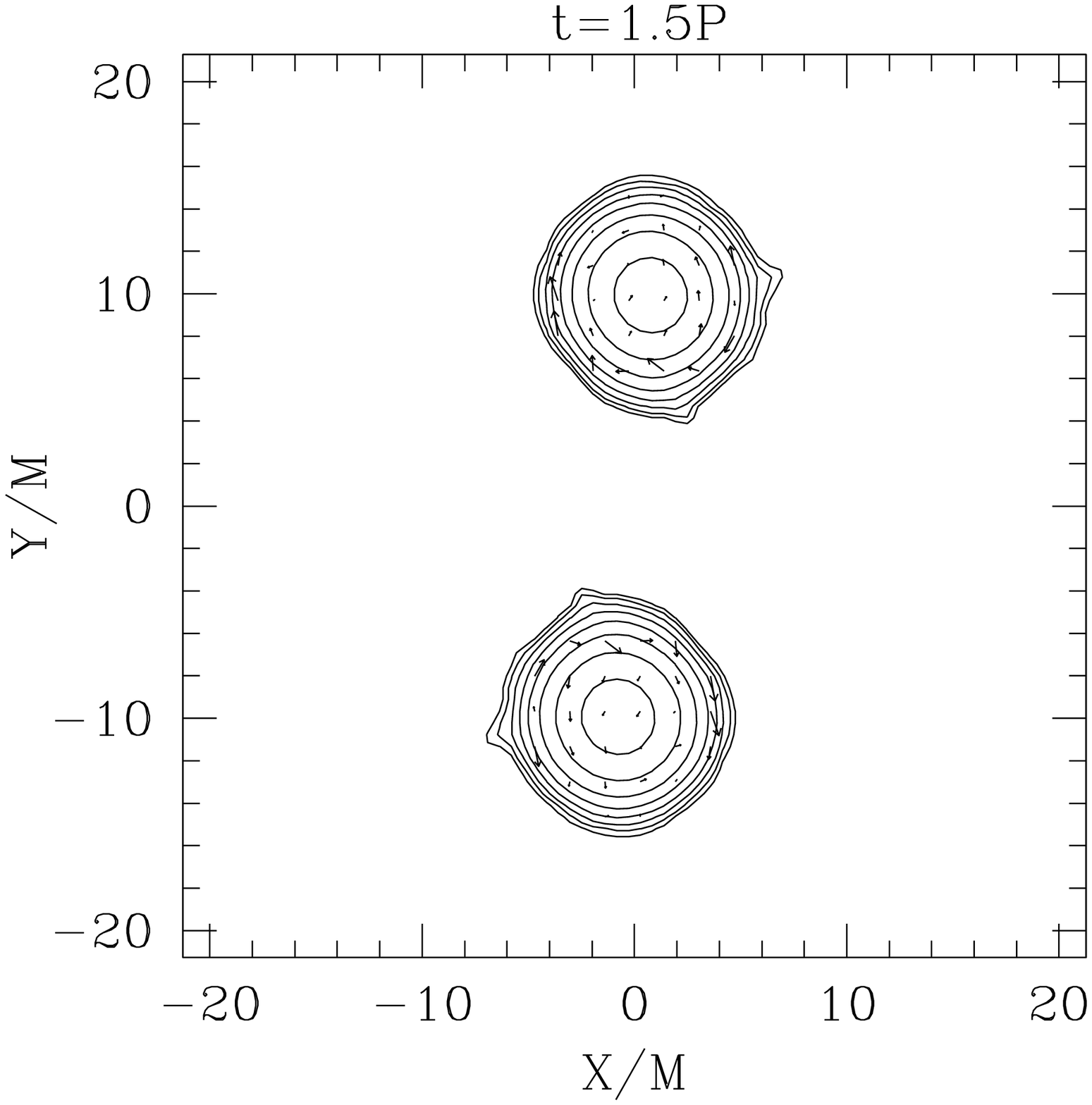} \\
\epsfxsize=2.6in
\leavevmode
\epsffile{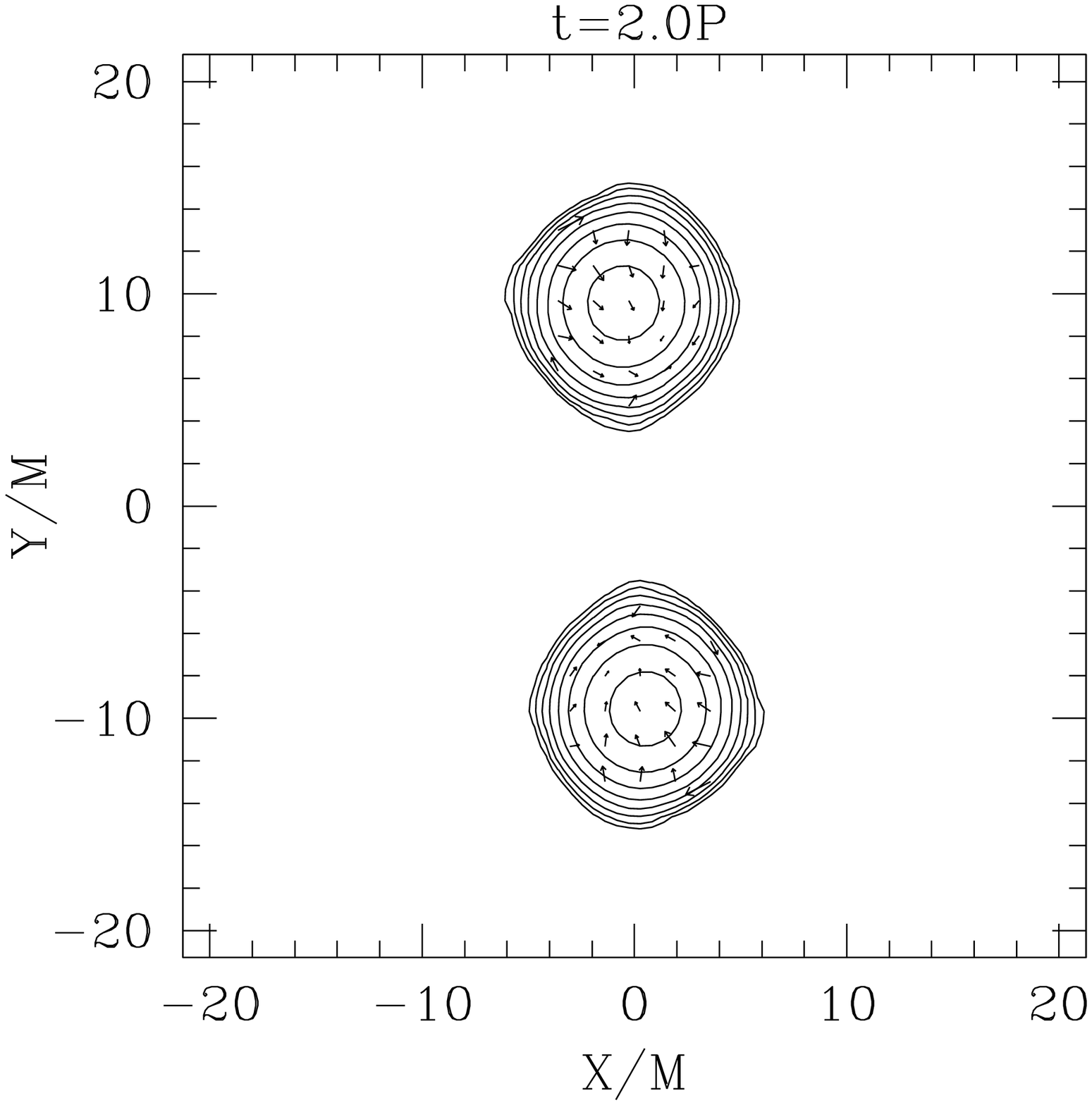}
\epsfxsize=2.6in
\leavevmode
\caption{
Snapshots in a rotating coordinate frame of the rest density contour lines and 
the velocity field in the equatorial plane
for a simulation of a corotating binary. The contour lines are drawn for 
$\rho_0=10^{-(0.2~j+0.1)} \rho_{0i}^{Max} $, where $\rho_{0i}^{Max}$ 
denotes the maximum value of the rest-density $\rho_0$ at $t=0$
(here it is $0.0573$), 
for $j=0,1,..,7$. Vectors indicate the local velocity field and the scale
is as shown in the top left-hand frame. The stars are orbiting clockwise
in the {\it inertial} reference frame with an initial coordinate velocity
of $0.102c$. $P$ denotes the initial orbital period.}
\label{bin_contour}
\end{center}
\end{figure*}


\begin{figure}
\epsfxsize=3.0in
\begin{center}
\leavevmode \epsffile{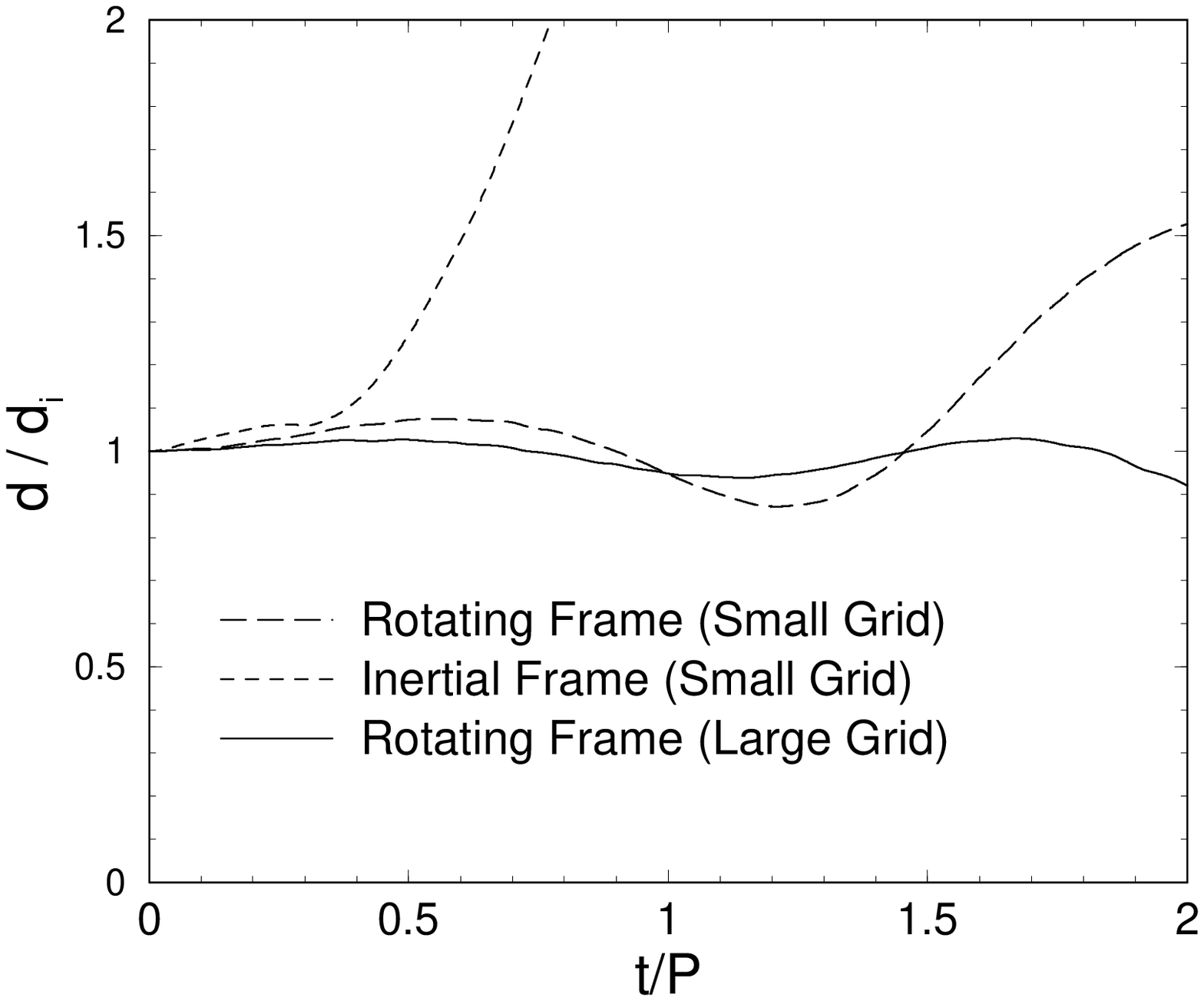}
\end{center}
\caption{ Evolution of the coordinate separation between the maximum
  rest mass density points of the two stars in the binary system shown in
  Fig.~\ref{bin_contour}.
  The time is given as a fraction of the initial orbital period
  and the separation $d$ as a fraction of the initial value $d_i$.}
\label{bin_sep}
\end{figure}

Binary neutron stars are among the most promising sources of
gravitational radiation for the new generation of gravitational wave
interferometers.  This makes the numerical simulation of such systems
one of the most important goals of a fully relativistic hydrodynamics
code and provides one of the most demanding tests for any such code.
A binary system allows us to uncover potential
problems that may not be evident in axisymmetric scenarios.
Previous simulations have focused on the coalescence and
merger of binary neutron stars \cite{Shibata:1999wm,Shibata:2002jb}.
In this Section we demonstrate that our code can stably evolve
binaries in stable, quasi-circular orbits for over two periods (compare
\cite{Shibata:1999aa}).

As initial data for these simulations we adopt the data of
\cite{bcsst97b,bcsst98b}, describing two equal mass polytropes in
co-rotating, quasi-circular orbit.  These data have been constructed
using the conformal ``thin-sandwich''
decomposition of the constraint equations
\cite{Wilson:1989,Wilson:1995ty,Wilson:1996ty,York:1999,Cook:2000}
together with maximal slicing and spatial conformal flatness.

In this Section we focus on one particular case and postpone a more
complete presentation for a forthcoming paper \cite{Marronetti:2002}.
We model the neutron stars as $\Gamma = 2$ polytropes with an
individual rest mass of $M_0^{\rm ind} = 0.1$ in our nondimensional
units (recall that the polytropic index $\kappa$ is set to unity).  At
infinite separation, this corresponds to an individual gravitational
mass of $M_{\infty}^{\rm ind}=0.096$.  The compaction of $(M^{\rm
ind}/R)_{\infty}=0.088$ implies that the gravitational fields are
moderately relativistic (the maximum compaction for $\Gamma = 2$
polytropes is $(M^{\rm ind}/R)_{\infty}=0.21$).  We adopt initial data
for a binary separation of $z_a=0.3$, where $z_a$ is the ratio between
the coordinate separation from the center of mass to the nearest point on the
star's surface to the farthest point (see \cite{bcsst97b,bcsst98b}), meaning
that the separation between the stellar surfaces is about $86\%$ of a
stellar diameter.  This separation is well outside the innermost
stable circular orbit (ISCO) as determined by the analysis of initial
data sets (see \cite{bcsst98a}).  At this separation, the total binary
ADM mass is $M = 0.19$ and the total
angular momentum is $J/M^2 = 1.36$.

We evolve these initial data on three different grids.  Two ``small
grid'' simulations are evolved on $120\times60^2$ gridpoints, with a
resolution of $\Delta x = \Delta y = \Delta z = 0.55 M$ (the binary is
symmetric across the equatorial plane, and, for equal mass stars,
$\pi$-rotation symmetric around the center of mass).  The individual
stars are resolved by $\approx 16$ gridpoints across the stellar
diameter (compare \cite{Shibata:2002jb} where much larger grids are
used).  One of these small grid evolutions is performed in the
inertial frame, the other in a rotating frame.  On these small
(uniform) grids, the outer boundaries are imposed very close to the
stars (at a separation of two stellar diameters), which we
found to introduce numerical noise.  We therefore repeated these 
simulations on a ``large grid'', performed in a rotating frame, where 
we doubled the number of gridpoints and the separation to the outer 
boundary, while keeping the grid resolution constant. 
This corresponds to a numerical grid covering a cubic 
coordinate volume of side [-66,66] in the units of Fig.
\ref{bin_contour}. The size of this grid is such that the corner 
points almost ``touch'' the surface of the light cylinder,
the cylinder with coordinate radius $R_L = 1/\Omega$, where $\Omega$
is the rotating frame angular frequency. The possibility
of evolution in rotating frames with grids that extend beyond the
light cylinder will be studied in a future article.

We used Courant factors of 0.30 and 0.46 for the small and 
large grid runs respectively, resulting in about 3,000
timesteps per orbit for the latter case.
We adopt the K-driver (\ref{K_driver}) for the lapse (using
$\epsilon=0.625$, $c=0.1$, and 5 substeps per step) and the
gamma-driver (\ref{gamma_driver}) for the shift (using $\eta=0.2$,
$k=0.005$ and 10 substeps). We used Sommerfeld type
boundary conditions for the gravitational fields (see Section
\ref{fields_BCs}) and {\it Copy} type for the hydrodynamical fields
(see Section \ref{hydro_BCs}).  For the artificial viscosity we used
$C_{\rm Qvis} = 0.1$ and $C_{\rm Lvis} = 0$ (see Section
\ref{art_vis}).  We also used $c_H = 0.050 \Delta T$ in
(\ref{phi_nevolve}).  For the small, rotating frame grid, this choice
led to the code crashing after about a period and a half.  However, we
found that restarting the code just before that with $c_H = 0.024
\Delta T$ allowed us to continue the evolution.  For the large grid,
no such adjustment was necessary.  For both small grid simulations we
found that the stars quickly drift apart.  To compensate for this we
reduced the orbital angular velocity $\Omega$ by 2\% for these two
cases.  Again, no such adjustment was required for the large grid.

In Fig.~\ref{bin_contour} we show contour plots of the rest density
$\rho_0$ at half period intervals for the large grid simulation in the
rotating frame.  The three-velocity of the fluid is represented by the
arrows.  The fact that the different panels look almost identical
indicates how well the binary remains in its circular orbit.

Imposing the outer boundaries at smaller separations we were unable to
keep the binary in circular orbit.  In Fig.~\ref{bin_sep} we plot the
coordinate separation $d$ between the two points of maximum density
for all three simulations as a function of time~\cite{note2}.  For the
small grid in the inertial frame, the two stars start to drift apart after 
a very short time.  When performing the same simulation in a rotating frame,
the stars remained in binary orbit for about two periods, but
ultimately merge.  This merger is triggered by the development of
orbital eccentricity. 
When we compare this small grid run with the large grid simulation, we see 
that the eccentricity is greatly reduced \cite{note3}. This result seems to
validate the quasi-equilibrium approach to obtaining reasonable initial 
data for corotating neutron star binaries in circular equilibrium and underlies 
the importance of the boundary proximity effect in these simulations.

\begin{figure}
\epsfxsize=3.0in
\begin{center}
\leavevmode \epsffile{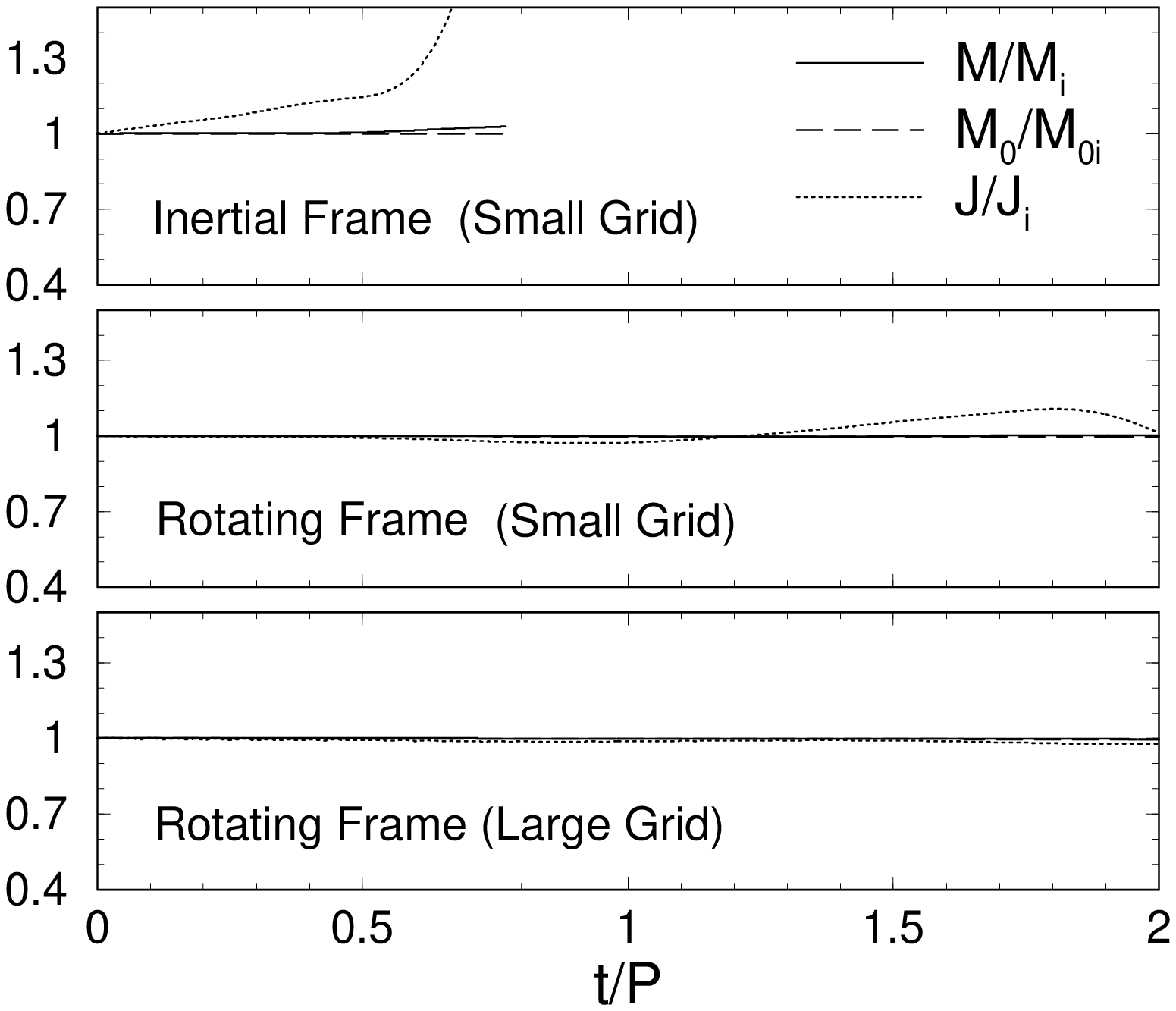}
\end{center}
\caption{Rest mass, gravitational mass, and angular momentum 
for the three simulations depicted in Fig.~\ref{bin_sep} .}  
\label{bin_ADM}
\end{figure}

\begin{figure}
\epsfxsize=3.0in
\begin{center}
\leavevmode \epsffile{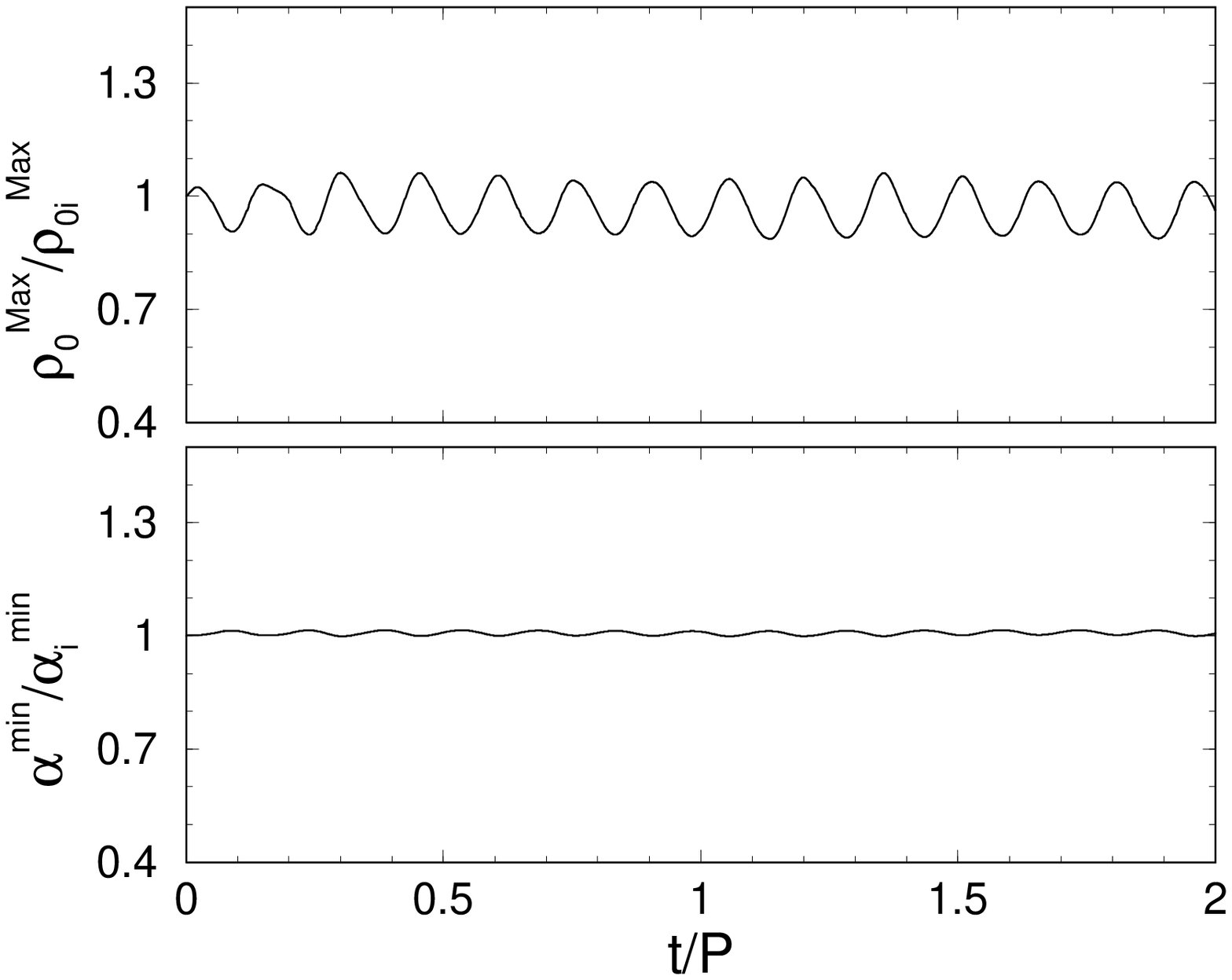}
\end{center}
\caption{Maximum rest mass density $\rho_0$ (top) and minimum lapse
function $\alpha$ (bottom) for the large grid, rotating frame
simulation.}
\label{bin_rho}
\end{figure}

\begin{figure}
\epsfxsize=3.0in
\begin{center}
\leavevmode \epsffile{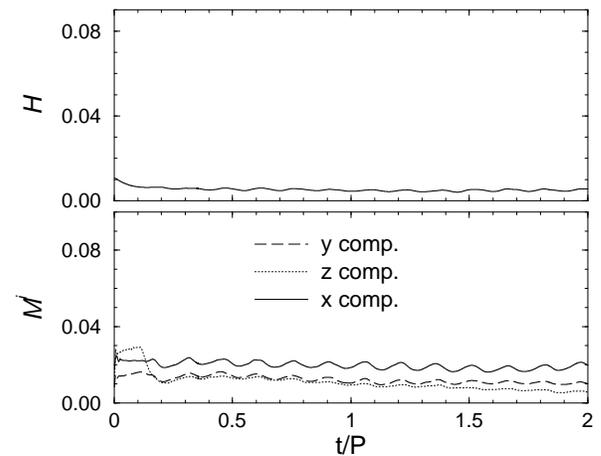}
\end{center}
\caption{L2 norms of the Hamiltonian constraint $\mathcal{H}$
(top) and momentum constraints $\mathcal{M}^i$ (bottom) 
for the large grid, rotating frame simulation.
All the curves have been normalized as explained in Sec. \ref{Bin}.}
\label{bin_constr}
\end{figure}

\begin{figure}
\epsfxsize=3.0in
\begin{center}
\leavevmode \epsffile{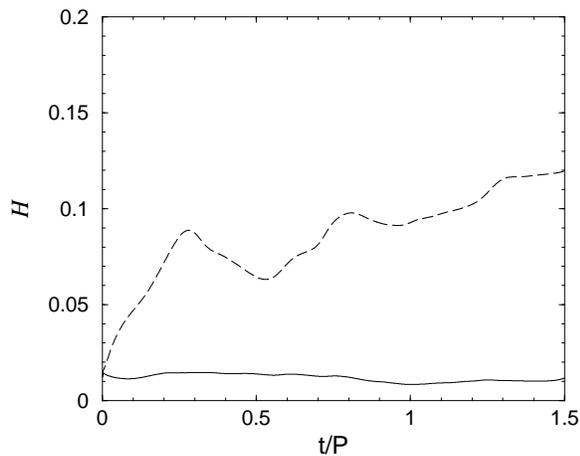}
\end{center}
\caption{L2 norms of the Hamiltonian constraint $\mathcal{H}$
for small grid, rotating frame simulations, using $c_H=0.05\Delta T$
(solid curve) and $c_H=0.00$ (dashed curve). This plot shows the
effect of the $c_H$ term in Eq. (\ref{phi_nevolve}) 
on the conservation of the Hamiltonian constraint.}
\label{bin_ch}
\end{figure}

\begin{figure}
\epsfxsize=3.0in
\begin{center}
\leavevmode \epsffile{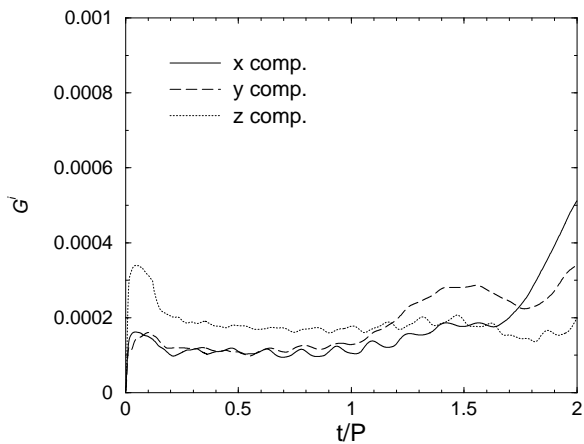}
\end{center}
\caption{L2 norm of the gamma constraint $\mathcal{G}^i$ for the large
grid, rotating frame simulation.}
\label{bin_Gamma}
\end{figure}

We show more diagnostics of these runs in Figs.~\ref{bin_ADM} through
\ref{bin_Gamma}.  In Fig.~\ref{bin_ADM} we show the rest mass $M_0$,
gravitational mass $M$, and angular momentum $J$ for the three
different simulations.  The use of a rotating
frame, which minimizes fluid advection through the numerical grid,
leads to large improvements, especially in the conservation of
angular momentum. Close outer boundaries lead to considerable 
numerical error. The 
system loses mass and angular momentum through the emission of
gravitational radiation, but at a rate that should lead to smaller
deviations than we find in our simulations. 
The maximum variation of the rest mass, gravitational mass, and 
angular momentum for the large grid run over the first two orbits 
was $0.3\%$, $0.3\%$, and $2.2\%$ respectively.

In Fig.~\ref{bin_rho} we show the maximum rest density $\rho_0$ and
the minimum value of the lapse $\alpha$ as functions of time.  The
small oscillations correspond to the fundamental mode of the
individual neutron stars, which are induced by the truncation error of
the finite grid resolution.  The period of the oscillations $P \sim
16.6$ agrees well with the theoretical value of $t=16.0$ (see Fig.~32
in \cite{Shibata:1999aa}). These oscillations are not damped, since
for these runs we switched off the linear viscosity terms ($C_{\rm
Lvis} = 0$).

We monitor the Hamiltonian (\ref{Hamiltonian_BSSN}), momentum
(\ref{momentum_BSSN}), and Gamma (\ref{gamma_BSSN}) constraints in
Figs.~\ref{bin_constr}-\ref{bin_Gamma}.  We show the L2 norms of
the corresponding constraint violations. For the Hamiltonian and
momentum constraint, these violations are normalized with respect to
$N_{HC}$ and $N_{MC}$ evaluated at $t=0$ (see Eqs.~(\ref{Hnorm}) and
(\ref{Mnorm})). In Figure \ref{bin_ch} we show the small-grid
rotating-frame result for the Hamiltonian constraint violation (solid
line), as well as the result from a similar small-grid evolution in
which we set $c_H = 0$ (dashed line)~\cite{note4}.  The difference
between these two lines illustrates the effect of the addition of this 
particular term in Eq.~(\ref{phi_nevolve}) on the conservation of the
Hamiltonian constraint.
The constraint deviations $\mathcal{G}^i$, which are
particularly sensitive to the choice of spatial gauge, remain well behaved 
during the first two orbits in these evolutions with the Gamma-driver.

As this example demonstrates, our code is able to stably evolve
binaries in stable, quasi-circular orbits for over two orbital periods. 
In a forthcoming paper \cite{Marronetti:2002} we will
use this code to systematically study binary sequences, both to
dynamically locate the ISCO and to test how accurately currently
available quasi-equilibrium initial data represent binaries in
quasi-circular orbit.


\section{Summary}
\label{summary}

We have tested our 3+1 relativistic hydrodynamics code
on a variety of problems.  We find that our current algorithm,
supplemented by driver
gauge conditions, is rather robust.  The grid resources required for
stable evolution and reasonable accuracy are modest. 
We accurately evolve shock tubes, spherical dust collapse,
and relativistic spherical polytropes.  We also evolved uniformly and
differentially rotating equilibrium polytropes, and maintained stable
configurations stationary for several rotational periods.  Two
applications carried out with our code are particularly significant.
First, we examined the collapse from large radius of a star with
significant spin to a Kerr black hole.  Second, we evolved stable
binary neutron stars for several orbits, maintaining quasi-circular
equilibrium.  The first application
indicates that we can study the effects of angular momentum on
gravitational collapse and on the resulting waveform, an effort already
initiated in \cite{n81,Shibata00}.  The second application indicates that
we can identify and evolve dynamically stable quasi-circular neutron star
binaries.  This ability can be used to locate the ISCO dynamically and
to follow the transition from an inspiral to a plunge trajectory.
In addition, dynamic simulation allows us to improve binary
initial data, for example by allowing initial ``junk'' gravitational
radiation to propagate away.  We also hope to compute detailed gravitational
waveforms form these binaries, refining the wavetrains reported in
\cite{dbs01,dbssu02}.

We note that several challenges remain to be addressed before there exists
a code capable of modeling all the gravitational wave sources of
current astrophysical interest.  One problem is the need to maintain
adequate grid coverage of
the collapsing star or inspiralling binary while still keeping the
outer boundaries sufficiently distant, i.e. the problem of dynamic range.
Adaptive mesh
techniques far more sophisticated than the crude rezoning used here may be
necessary.  A related problem concerns gravitational wave
extraction, as it currently is not possible to place outer boundaries in
the wave zone.  Finally, the formation of black hole singularities in
hydrodynamic collapse scenarios remains an additional challenge to
determining the late-term behavior of such systems.  Special
singularity-handling techniques, such as excision, need to be developed
further.


\acknowledgments

It is a pleasure to thank Masaru Shibata and Hwei-Jang Yo for useful
discussions.  Most of the calculations were performed
at the National Center for Supercomputing Applications at the
University of Illinois at Urbana-Champaign (UIUC).  This paper was
supported in part by NSF Grants PHY-0090310 and PHY-0205155 and
NASA Grant NAG 5-10781 at UIUC and NSF Grant PHY 01-39907 at Bowdoin College.


\begin{appendix}

\section{Treatment of Advective Terms in the Hydrodynamic Equations}
\label{advection_appendix}

Solving Eqs.~(\ref{evolve_rhostar})-(\ref{evolve_stilde}) requires solving
equations of the form
\begin{equation}
\label{advection_eq}
{\partial q\over\partial t} + {\partial (q v)\over\partial x} = S
\end{equation}
Let us finite difference this equation.  Let $q^n_i$ be the value of
$q$ at gridpoint $i$ on time level $n$.  Let $\Delta x$ be the coordinate
distance between neighboring gridpoints, and $\Delta T$ be the
timestep.  The Courant factor is $C = \Delta T/\Delta x$.  Then we
difference (\ref{advection_eq}), say for the predictor step, as
\begin{equation}
\label{finite_difference}
q^{n+1}_i = q^n_i + \Delta T[(v^n_{i-1/2}q^n_{i-1/2} - v^n_{i+1/2}q^n_{i+1/2})
  + S^n_i]
\end{equation}
where
\begin{eqnarray}
\label{v_iface}
\displaystyle v^n_{i+1/2} &=& {\rho^n_{\star i}v^n_{i}
  + \rho^n_{\star i+1}v^n_{i+1}
  \over \rho^n_{\star i} + \rho^n_{\star i+1}} \\
\label{q_iface}
q^n_{i+1/2} &=& \left\{ \begin{array}{ll}
                        q^n_i + \Delta x \nabla^n_iq / 2
			    & \mbox{if $v^n_i > 0$} \\
			q^n_{i+1} - \Delta x \nabla^n_{i+1}q / 2
			    & \mbox{if $v^n_i < 0$}
			\end{array}
                 \right.
\end{eqnarray}
In (\ref{q_iface}),
\begin{equation}
\label{nablaq}
\nabla^n_iq = \left\{ \begin{array}{ll} \displaystyle 
               {2\Delta^n_{i-1/2}q\Delta^n_{i+1/2}q
               \over \Delta^n_{i-1/2}q + \Delta^n_{i+1/2}q}
			    & \mbox{if $\Delta^n_{i-1/2}q\Delta^n_{i+1/2}q 
                              > 0$} \\
               0 & \mbox{otherwise}
             \end{array}
           \right.
\end{equation}
where $\Delta^n_{i+1/2}q = (q^n_{i+1}-q^n_i)/\Delta x$.  In many van Leer
schemes, the term $q + \Delta x \nabla q / 2$ in Eq.~(\ref{q_iface}) is
replaced by the time-averaging expression
$q + (\Delta x - v\Delta T)\nabla q / 2$.  Since we use a
predictor-corrector method, we don't need to time average.


\section{Newtonian Limit of the Euler Equation in a Rotating Frame}
\label{newt_limit}

Here we recover the Newtonian Euler equation in a rotating frame, as 
given for instance in \cite{Tassoul}, taking the weak-field limit of
the general relativistic Euler equation.
All variables and differential operators are given in the rotating frame,
with the exception of ``barred'' quantities that reside in the inertial frame.
For simplicity, we will work with the general relativistic Euler equation 
written as a function of the variable $\hat{u}_k$, defined as
\begin{eqnarray}
\hat{u}_k \equiv h u_k = \hat{\gamma}_{ki} u^0 h e^{4 \phi} (v^i+\beta^i)\ .
\label{ulk}
\end{eqnarray}
Accordingly, we have
\begin{eqnarray}
 & & \partial_{t}(\rho_{\star} \hat{u}_k) + \partial_{i} 
	(\rho_{\star} \hat{u}_k v^i)
	=  \nonumber\\ 
 & & \qquad - \alpha e^{6 \phi} \partial_{k} P 
      - \rho_{\star} u^0 \alpha h \partial_{k} \alpha 
      - \rho_{\star} \hat{u}_j \partial_{k} \beta^j	\nonumber\\ 
 & & \qquad -  2 {\rho_{\star} h ((u^0 \alpha)^2-1) \over u^0 } 
	\partial_{k} \phi  
      + {\rho_{\star} e^{-4 \phi} \hat{u}_i \hat{u}_j \over 2 u^0 h} 
		\partial_{k} \hat{\gamma}^{ij} ~,\qquad \left.\right.
\label{greuler}
\end{eqnarray}
where $\rho_{\star}$ and $h$ are defined in Sec. (\ref{hydro_equations}). 
Using the continuity equation (\ref{continuity}) we can re-write the lhs of 
equation (\ref{greuler}) as
\begin{eqnarray}
\partial_{t} (\rho_{\star} \hat{u}_k) + \partial_{i} (\rho_{\star} \hat{u}_k v^i)
	= \rho_{\star} ( \partial_{t} \hat{u}_k + v^i \partial_{i} \hat{u}_k )~.
\label{eq0}
\end{eqnarray}
To obtain the Newtonian limit, we expand the different terms
in equation (\ref{greuler}) to first order in the Newtonian potential
$\phi_N$ and the square of the fluid velocity $v$. For instance:
\begin{eqnarray}
\hat{\gamma}_{ki} & \rightarrow & \delta_{ki} 		\nonumber\\
u^0 & \rightarrow & 1+{v^2 \over 2} 			\nonumber\\
h & \rightarrow & 1 					\nonumber\\
e^{\phi} & \rightarrow & 1-2\phi_N 			\nonumber\\
\alpha & \rightarrow & 1 + \phi_N			\nonumber\\
\beta^i = \bar{\beta}^i + ( \vec{\Omega} \times
	\vec{r} )^i & \rightarrow & ( \vec{\Omega} \times \vec{r} )^i ~,
\label{limits_1}
\end{eqnarray}
where $\bar{\beta}^i$ is the shift vector in a inertial frame, 
$\vec{\Omega}$ is the angular velocity of the rotating frame
with respect to the inertial frame, and $\vec{r} \equiv (x,y,z)$ is the 
position vector. The limits (\ref{limits_1}) combined with equation 
(\ref{ulk}) give
\begin{eqnarray}
\hat{u}_k \rightarrow v^k +  ( \vec{\Omega} \times \vec{r} )^k
\label{ulkN}
\end{eqnarray}

We proceed now to take the Newtonian limit of the rhs of Eq.~(\ref{eq0}). 
To do so, we note that
\begin{eqnarray}
\partial_{t} \hat{u}_k \rightarrow \partial_{t} v^k 
	+ \partial_{t} ( \vec{\Omega} \times \vec{r} )^k  = \partial_{t} v^k
\end{eqnarray}
since $\partial_{t} \vec{\Omega} = \partial_{t} \vec{r} = 0$, and 
\begin{eqnarray}
v^i \partial_{i} \hat{u}_k & \rightarrow &		
	v^i \partial_{i} 
	\left( v^k + ( \vec{\Omega} \times \vec{r} )^k \right) \nonumber\\
	& = & v^i \partial_{i} v^k + ( \vec{\Omega} \times \vec{v} )^k ~.
\end{eqnarray}
These conditions together with equation give (\ref{eq0}) 
give the Newtonian limit of the lhs of equation (\ref{greuler})
\begin{eqnarray}
 \partial_{t} (\rho_{\star} \hat{u}_k) & + & \partial_{i} 
	(\rho_{\star} \hat{u}_k v^i) \nonumber\\
 & \rightarrow & \rho_{\star} \left( \partial_{t} v^k + v^i \partial_{i} v^k 
	+ (\vec{\Omega} \times \vec{v})^k \right)~.
\label{eq0_lhs}
\end{eqnarray}

To work on the rhs of equation (\ref{greuler}) to Newtonian order,
we derive the following limits using Eqs. (\ref{limits_1}),
again keeping only the first order terms in $\phi_N$ and $v^2$:
\begin{eqnarray}
\alpha e^{6 \phi} \partial_{k} P & \rightarrow &  \partial_{k} P  \nonumber\\
\rho_{\star} u^0 h \alpha \partial_{k} \alpha & \rightarrow &  
				  \rho_{\star} \partial_{k} \phi_N \nonumber\\
2 \rho_{\star} h { ((u^0 \alpha)^2-1) \over u^0 } 
                    \partial_{k} \phi & \rightarrow & 0 
	 \nonumber\\
{\rho_{\star} e^{-4 \phi} \over 2 u^0 h } \hat{u}_i \hat{u}_j 
	\partial_{k} \hat{\gamma}^{ij} & \rightarrow & 0  ~.
\label{eq3}
\end{eqnarray}
The final term is composed using (\ref{ulkN}):
\begin{eqnarray}
\rho_{\star} \hat{u}_j \partial_{k} \beta^j \rightarrow 
	\rho_{\star} \left( v^j + ( \vec{\Omega} \times \vec{r} )^j \right) 
	\partial_{k} ( \vec{\Omega} \times \vec{r} )^j~.
\end{eqnarray}
We rewrite the first term above as
\begin{eqnarray}
\rho_{\star} v^j \partial_{k} ( \vec{\Omega} \times \vec{r} )^j & = &
	\rho_{\star} v^j ( \vec{\Omega} \times \partial_{k} \vec{r} )^j 
	    \nonumber\\
	& = & \rho_{\star} \epsilon_{kjn} v^j \Omega_n	   \nonumber\\
	& = & - \rho_{\star} ( \vec{\Omega} \times \vec{v} )^k  ~,
\label{eq4}
\end{eqnarray}
and the second term as
\begin{eqnarray}
\rho_{\star} ( \vec{\Omega} \times \vec{r} )^j 
                    \partial_{k} ( \vec{\Omega} \times \vec{r} )^j
  = - \rho_{\star} \left( \vec{\Omega} \times 
                            (\vec{\Omega} \times \vec{r}) \right)^k ~.
\label{eq5}
\end{eqnarray}

Combining (\ref{eq0_lhs}), (\ref{eq3}), (\ref{eq4}), and (\ref{eq5}), 
yields
\begin{eqnarray}
& &\rho_{\star}\left( \partial_{t} v^k + v^i \partial_{i} v^k 
	+ (\vec{\Omega} \times \vec{v})^k \right)  = \nonumber \\
& &\qquad\qquad -\partial_{k} P - \rho_{\star} \partial_{k} \phi_N 
- \rho_{\star} (\vec{\Omega} \times \vec{v})^k \\
 &  &\qquad\qquad -\rho_{\star} \left( \vec{\Omega} 
       \times (\vec{\Omega} \times \vec{r}) \right)^k~.
\nonumber
\end{eqnarray}
Rearranging terms and replacing $\rho_\star$ by its limit the mass density
$\rho$ yields the Newtonian limit of the general relativistic Euler
equation 
(\ref{greuler}) :
\begin{eqnarray}
\partial_{t} v^k + v^i \partial_{i} v^k & = &\nonumber \\
- {1 \over \rho} \partial_{k} P - \partial_{k} 
        \phi_N & - & 2 (\vec{\Omega} \times \vec{v})^k
-\left( \vec{\Omega} \times (\vec{\Omega} \times \vec{r}) \right)^k ~,
\qquad \left.\right.
\end{eqnarray}
where the last two terms of the rhs correspond to the familiar
Coriolis and centrifugal force terms.


\section{ADM Mass and Angular Momentum in Rotating Frames}
\label{ADM_rotating}

In this Appendix, the ``barred'' fields represent variables in the inertial
frame, while the non-barred ones are quantities in rotating frames. In Sec.
\ref{diag}, we defined the total mass and angular momentum of an
asymptotically flat spacetime by two surface
integrals (Eqs. (\ref{Mdef}) and (\ref{Jdef}) respectively) which
characterize the asymptotic behavior of the metric on a time slice. 
These surface integrals
were transformed into the volume integrals (\ref{Mfin}) and (\ref{Jfin})
according to the calculation described in~\cite{yo02}.  These volume
integrals are numerically evaluated in our code on the computational grid. 
When working in rotating frames, one might worry that these integrals do
not apply, since the 4-metric is not asymptotically flat due to the
$\vec{\Omega} \times \vec{r}$ term in the shift.  It turns out that this
is not a problem, since the surface integral formulae for $M$ and
$J$ can be obtained assuming only that the 3-metric and extrinsic
curvature are asymptotically flat~\cite{myby}.  Therefore, we can
evaluate the volume integrals (\ref{Mfin}) and (\ref{Jfin}) in the
rotating frame and be sure that the $M$ and $J$ that we find at a
given time will be the same as what we would have found by transforming
into an inertial frame and then computing the integrals.  We can see this
explicitly by transforming the integrands from an inertial to a
rotating frame.  For example, the mass (\ref{Mfin}) written in terms of
the ``barred'' inertial frame quantities is
\begin{eqnarray}
\label{MfinApp}
M   &=& \int_V \Bigl(e^{5\bar{\phi}}(\bar{\rho} 
   + {1\over 16\pi}\bar{\tilde A}_{ij} \bar{\tilde A}^{ij}
   - {1\over 24\pi}\bar{K}^2) \\
    & &  \quad - {1\over 16\pi}\bar{\tilde\Gamma}^{ijk}
	\bar{\tilde\Gamma}_{jik}
	+ {1-e^{\bar{\phi}}\over 16\pi}\bar{\tilde R}\Bigr) d^3\bar{x}~~.
			     \nonumber 
\end{eqnarray}
For simplicity, we take the inertial coordinate
system to be the one which is instantaneously aligned with our rotating
frame at the time that we are computing $M$ and $J$.  Then the
transformation is given by Eqs. (\ref{rot_fields}), (\ref{rot_matter}), and
(\ref{spatial_tensor}).  (From Eq.~(\ref{rho_def}), we see that $\rho$ is
an invariant.)  Applying these rules, we see that every term in the
integrand is identical in the inertial and rotating frames.  The same is
true of the integrands for $J$ and $M_0$.

\end{appendix}


\begin{thebibliography}{99}
\bibitem{note1} We actually use
$\tilde S_k \equiv h S_k$ as our variable, where $h$ is the enthalpy. In
this section, we will ignore this point for the sake of simplicity.
\bibitem{note2} We define the coordinate separation by
  $d = (\int_V d^3x f(\rho_{\star}r)) /
                     ( \int_V d^3x f(\rho_{\star}))$,
  where $f(x) = x$ if $\rho_0$ is greater than 0.01
  times the maximum value of $\rho_0$ on the grid, and $f(x) = 0$
  otherwise.
\bibitem{note4} This comparison run was performed in a small grid due to the
  huge computational resources consumed by large grid runs. However, the
  difference between the two curves is similar to the one obtained when
  comparing two small grid runs.
\end{thebibliography}
\end{document}